\documentclass[%
 reprint,
 superscriptaddress,
 amsmath,amssymb,
 aps,
]{revtex4-2}

\usepackage{prmagnet}

\myexternaldocument{suppinfo}{si-}

\begin{document}

\title{ Kibble–Zurek dynamics beyond the thermodynamic limit }

\author{Max Haas}
\thanks{These authors contributed equally to this work.}
\affiliation{Department of Materials Science and Engineering, Norwegian University of Science and Technology (NTNU), 7034, Trondheim, Norway}
\affiliation{German Aerospace Center (DLR), Institute of Materials Research, 51147, Cologne, Germany}

\author{Guy C. Moore}
\thanks{These authors contributed equally to this work.}
\affiliation{Materials Sciences Division, Lawrence Berkeley National Laboratory, CA 94720 Berkeley, California, USA}
\affiliation{Molecular Foundry, Lawrence Berkeley National Laboratory, Berkeley, CA 94720, USA}

\author{Isaac M. Craig}
\affiliation{Materials Sciences Division, Lawrence Berkeley National Laboratory, CA 94720 Berkeley, California, USA}
\affiliation{Molecular Foundry, Lawrence Berkeley National Laboratory, Berkeley, CA 94720, USA}

\author{Kristin A. Persson}
\affiliation{Materials Sciences Division, Lawrence Berkeley National Laboratory, CA 94720 Berkeley, California, USA}
\affiliation{Molecular Foundry, Lawrence Berkeley National Laboratory, Berkeley, CA 94720, USA}
\affiliation{Department of Materials Science and Engineering, University of California Berkeley, Berkeley, CA 94720, USA}

\author{Jiali He}
\affiliation{Department of Materials Science and Engineering, Norwegian University of Science and Technology (NTNU), 7034, Trondheim, Norway}

\author{Michael E. Hjelmstad}
\affiliation{Oxford Instruments Americas, Pleasanton, CA 94566, USA}

\author{Dennis Meier}
\email{dennis.meier@uni-due.de}
\affiliation{Department of Materials Science and Engineering, Norwegian University of Science and Technology (NTNU), 7034, Trondheim, Norway}
\affiliation{Faculty of Physics and Center for Nanointegration Duisburg-Essen (CENIDE), University of Duisburg-Essen, Duisburg, 47057, Germany}
\affiliation{Research Center Future Energy Materials and Systems, Research Alliance Ruhr, 44801, Bochum, Germany}

\author{Sinéad M. Griffin}
\email{sgriffin@lbl.gov}
\affiliation{Materials Sciences Division, Lawrence Berkeley National Laboratory, CA 94720 Berkeley, California, USA}
\affiliation{Molecular Foundry, Lawrence Berkeley National Laboratory, Berkeley, CA 94720, USA}

\author{Jan Schultheiß}
\email{jan.schultheiss@smn.uio.no}
\affiliation{Department of Materials Science and Engineering, Norwegian University of Science and Technology (NTNU), 7034, Trondheim, Norway}
\affiliation{Department of Mechanical Engineering, University of Canterbury, 8140, Christchurch, New Zealand}
\affiliation{Centre for Materials Science and Nanotechnology (SMN), University of Oslo, 0371, Oslo, Norway}

\date{\today}

\begin{abstract}
The Kibble–Zurek mechanism (KZM) predicts the universal scaling of topological defect densities generated during continuous phase transitions, a result derived in the thermodynamic limit of homogeneous systems. Yet real materials are intrinsically finite and heterogeneous, raising the  question of how nonequilibrium universality survives outside this idealized limit. Here we establish an ensemble-finite-size extension of the KZM that describes critical dynamics in collections of independent finite subsystems with distributed sizes. We test this framework using a polycrystalline model system undergoing a symmetry-breaking transition. We find that the density of topological vortices follows an apparent power law with an exponent approximately half that observed in single crystals. This reduced exponent emerges quantitatively from averaging over the experimentally measured grain-size distribution, without fitting to the measured vortex densities. Our results demonstrate that finite-size effects do not simply suppress KZ scaling but reorganize it into an emergent statistical scaling regime governed by system-size heterogeneity. This establishes microstructure and geometry as control parameters of nonequilibrium critical dynamics and provides a general strategy to engineer scaling behaviour in complex materials.
\end{abstract}

\maketitle

The Kibble–Zurek mechanism (KZM) provides a universal framework for nonequilibrium critical dynamics: as a system is driven through a continuous phase transition, the density of topological defects is determined by the quench rate~\cite{Kibble1976,zurek1985cosmological}. 
In its canonical form the KZ scaling law reads 
\begin{equation}
n \propto R^{\mu}, \qquad  
\mu = \frac{\nu d}{1 + z\nu},
\label{eq:KZMexponent}
\end{equation}
where $n$ is the topological defect density, $R$ is the quench rate, $\mu$ is the KZ scaling exponent, $\nu$ and $z$ are the equilibrium critical exponents, and $d$ is the dimensionality. 
The KZM has been experimentally demonstrated in diverse classical and quantum systems, including liquid crystals~\cite{Chuang1991,Ducci1999}, superfluid helium~\cite{Hendry1994,Bauerle1996}, superconductors~\cite{Carmi2000}, and trapped-ion crystals~\cite{Ulm2013}, establishing it as a cornerstone of symmetry-breaking dynamics.

In the standard formulation, the KZM assumes the thermodynamic limit: a homogeneous medium large enough that the correlation length $\xi$ and relaxation time $\tau_{\mathrm{eq}}$ diverge as $T\!\to\!T_c$. 
Critical slowing down causes the system to fall out of equilibrium when the rate of change of the control parameter exceeds $\tau_{\mathrm{eq}}^{-1}$, freezing correlations of characteristic size $\hat{\xi}\!\sim\!R^{-\nu/(1+z\nu)}$ and producing a defect density $n\!\sim\!\hat{\xi}^{-d}$. This description presumes a single characteristic length scale and neglects explicit cutoffs introduced by finite geometry, disorder, or heterogeneity.

While the KZM has been established in idealized systems approaching the thermodynamic limit, real materials are intrinsically finite and heterogeneous. Microstructural boundaries, such as grain interfaces, impose explicit length-scale cut-off points that can truncate the growth of correlations during a quench, 
modifying the dynamics that govern KZ scaling. 
These finite-size and finite-time effects smooth the divergences expected at a continuous transition, transforming sharp singularities into grain-size-dependent maxima. A wide range of theoretical studies—spanning generalized finite-time and finite-size scaling~\cite{liu2014dynamic,chandran2012kibble,polkovnikov2005universal,deng2009dynamical,de2011universal,kolodrubetz2012nonequilibrium,liu2013quasi,de2013microscopic,zhong2005dynamic,gong2010finite}, Monte Carlo simulations of XY and Ising models~\cite{jelicQuenchDynamics2d2011,liu2014dynamic}, as well as field-theoretic and Langevin analyses~\cite{rojasDynamicalScalingTwodimensional1999,yurkeCoarseningDynamicsXY1993}—have examined how confinement alters KZ dynamics. While these works show sensitivity to boundary conditions and dimensionality, they remain largely limited to idealized or low-dimensional models.  
How such finite-size effects manifest in real materials, where microstructure and boundaries are intrinsic, remains unexplored.

Here we address this with an extension of the KZM that captures nonequilibrium critical dynamics in ensembles of finite, weakly coupled subsystems. 
In this ensemble-finite-size KZM (EFS–KZ) framework, each finite subsystem of characteristic dimension $L$ obeys Dirichlet boundary conditions that truncate the correlation length $\xi\!\le\!L$.  
Averaging these truncated correlations over the distribution of sub-system sizes $f(L)$ yields an effective scaling law that interpolates smoothly between the thermodynamic and confined limits.  
This ensemble treatment predicts a systematic reduction of the apparent scaling exponent, $\mu_{\mathrm{eff}} < \mu_{\mathrm{bulk}}$, with the degree of renormalization determined by the distribution of subsystem sizes. We test these predictions experimentally in polycrystalline hexagonal ErMnO$_3$, whose microstructure provides a natural platform to probe this statistical finite-size regime.  Our framework establishes a bridge between idealized KZ scaling and the complex microstructures of real materials, unifying finite-size, finite-time, and disordered dynamics under a single theory.

\section*{Ensemble-finite-size KZM framework}
\begin{figure*}[t]
\centering
\includegraphics[height=0.345\textwidth]{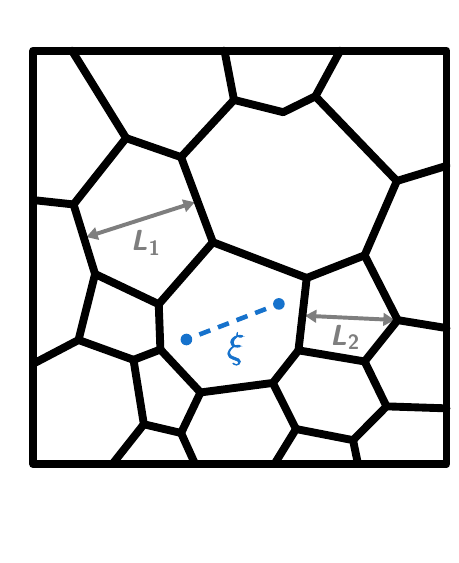}%
\makebox[0pt][r]{\raisebox{0.305\textwidth}{{\fontfamily{phv}\selectfont\bfseries\large a}\hspace{0.2685\textwidth}}}\hfill
\includegraphics[height=0.345\textwidth]{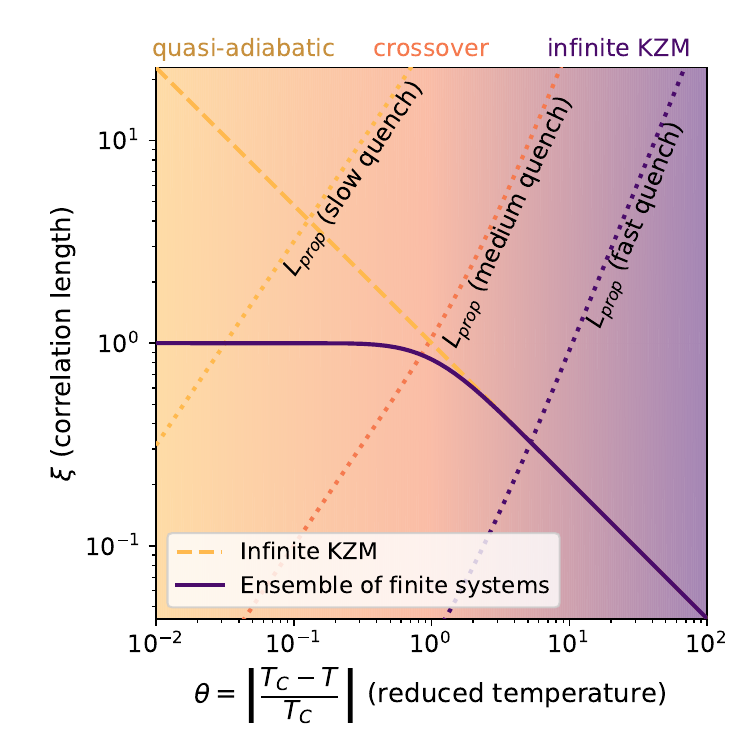}%
\makebox[0pt][r]{\raisebox{0.305\textwidth}{{\fontfamily{phv}\selectfont\bfseries\large b}\hspace{0.3094\textwidth}}}\hfill
\includegraphics[height=0.345\textwidth]{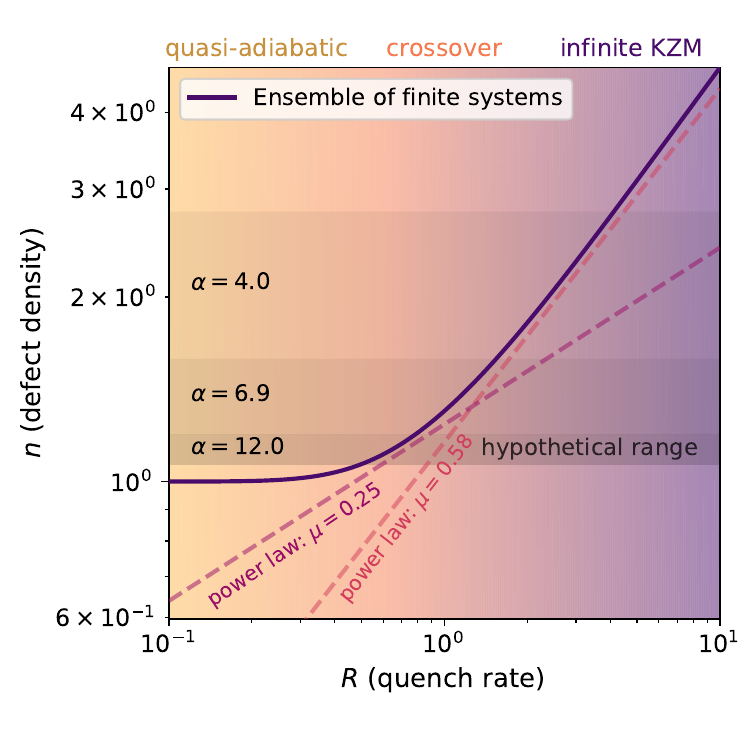}%
\makebox[0pt][r]{\raisebox{0.305\textwidth}{{\fontfamily{phv}\selectfont\bfseries\large c}\hspace{0.3238\textwidth}}}
\caption{
\textbf{Ensemble-finite-size Kibble–Zurek (EFS–KZ) scaling.}
\textbf{(a)}~Schematic of the competing length scales: the intrinsic KZ freeze-out length $\hat{\xi}_R=\xi_0R^{-\nu/(1+z\nu)}$ (blue dashed line) and the finite subsystem size~$L$ (grains in our test system). Subsystems with $L>\hat{\xi}_R$ follow bulk KZ scaling, whereas smaller subsystems are truncated at their boundaries. 
\textbf{(b)}~Correlation length~$\xi$ versus reduced temperature $\theta=|T_c-T|/T_c$ for the infinite-system KZM (orange dashed) and the EFS–KZ model (purple). The divergence of~$\xi$ is suppressed once $\xi>L$, producing a crossover from quasi-adiabatic to finite-size-limited dynamics as indicated by the shaded regions. Dotted lines mark the information propagation length $L_{\text{prop}}$ for slow, medium, and fast quenches. 
\textbf{(c)}~Predicted defect density~$n$ versus quench rate~$R$, illustrating the crossover between the bulk KZ regime ($n\!\propto\!R^{\mu_{\mathrm{bulk}}}$) and the finite-size-saturated regime. Dashed lines show power-law slopes $\mu_{\mathrm{bulk}}=0.58$ and $\mu_{\mathrm{eff}}=0.25$. Shaded bands illustrate the dependence of the absolute density scale on the nonuniversal prefactor $\alpha=4.0$, $6.9$, and $12.0$, which relates the freeze-out correlation length to the defect density ($n \propto \alpha^d \bar{\xi}_f^{-d}$, see Supplementary Sec.~\refSI{sec:parameters}{S2}). Changing $\alpha$ shifts the curves vertically without altering the crossover shape or apparent exponent. To facilitate comparison across scales, $n$, $R$, and related quantities are expressed in dimensionless form following the normalization introduced in Supplementary Sec.~\refSI{sec:KZfinite}{S1}.
}
\label{fig:theory}
\end{figure*}

The intrinsic spatial inhomogeneity in polycrystalline materials can be viewed as ensembles of finite subsystems with distributed characteristic sizes. This motivates an ensemble formulation of nonequilibrium critical dynamics, in which the emergent behaviour arises from statistical averaging over many finite, weakly coupled subsystems. We model a heterogeneous system as an ensemble of $N$ subsystems, each with characteristic dimension $L_i$ and independent KZ dynamics. Within each subsystem, the correlation length $\xi$ cannot exceed the geometric cutoff $L_i$, yielding a truncated correlation length
\begin{equation}
\xi_{L_i}(\theta) \approx \min(\xi_\infty(\theta), L_i)
\label{eq:xi_i}
\end{equation}
%
%
where $\xi_\infty(\theta) = \xi_0\,\theta^{-\nu}$ is the bulk correlation length as a function of reduced temperature $\theta = (T_c - T)/T_c$, and $\xi_0$ is the microscopic correlation length. Two characteristic length scales thus compete: the diverging bulk correlation length $\xi_\infty(\theta)$ and the finite subsystem size $L_i$. Subsystems with $L_i > \xi_\infty$ follow bulk critical scaling, whereas those with $L_i < \xi_\infty$ have their correlations truncated at their boundaries (Fig.~\ref{fig:theory}a and Supplementary Sec.~\refSI{sec:kz_subtle}{S1 A}).

In the limit of an infinite number of subsystems, \mbox{$N \rightarrow \infty$}, the effective correlation length of the full ensemble is obtained by an integral average over the distribution of subsystem sizes $f(L)$:
\begin{equation}
\tilde{\xi}(\theta) = \int_0^\infty \xi_{L}(\theta)\,f(L)\,dL
\label{eq:eff_corr_length_main}
\end{equation}
%
%
(see Supplementary Sec.~\refSI{sec:math_details}{S1 B} for a detailed derivation). In practice, $f(L)$ may represent the distribution of grain sizes in a polycrystal, structural widths, or other microstructural length scales.

For analytical treatment we approximate $f(L)$ by a log-normal form,
\begin{equation}
f(L)=\frac{1}{L\,\sigma_{\ln L}\sqrt{2\pi}}
\exp\!\left[-\frac{(\ln L-\langle \ln L \rangle)^2}{2\sigma_{\ln L}^2}\right],
\label{eq:lognormal}
\end{equation}
where $\langle \ln L \rangle$ and $\sigma_{\ln L}$ are the mean and standard deviation of $\ln L$, consistent with  studies of grain-size distributions~\cite{bergmannFormationPolycrystallineSilicon1998}.

The frozen-in correlation length that determines the defect density is obtained by solving for the freeze-out temperature $\theta_f$ at which the system can no longer evolve adiabatically. Following the formulation of Ref.~\cite{Griffin2012}, we determine $\theta_f$ from the self-consistent condition that the correlation length equals the integrated information propagation length $L_{\text{prop}}$,
\begin{equation}
\tilde{\xi}(\theta_f) = L_{\text{prop}}(\theta_f) = \frac{T_c}{R}\int_0^{\theta_f} d\theta\;\frac{\tilde{\xi}(\theta)}{\tilde{\tau}(\theta)},
\label{eq:freeze_out}
\end{equation}
where $R$ is the quench rate for a linear temperature ramp $T = T_c - R\,t$.  We assume that the effective relaxation time retains the dynamical scaling relation $\tilde{\tau}(\theta) = \tilde{\xi}(\theta)^z$, consistent with prior theoretical work on finite-size dynamics~\cite{luscombeFinitesizeScalingGlauber1996}. The defect density is given by $n = \alpha^d \tilde{\xi}(\theta_f)^{-d}$, where $d$ is the spatial dimensionality and $\alpha$ is a dimensionless scale prefactor relating the freeze-out correlation length to the physical inter-defect spacing (see Supplementary Sec.~\refSI{sec:parameters}{S2}). 

The key consequence of Eq.~(\ref{eq:eff_corr_length_main}) is that the effective correlation length $\tilde{\xi}(\theta)$ interpolates smoothly between two regimes as a function of $\theta$. Far from criticality, where $\xi_\infty(\theta) \ll \langle L \rangle$, virtually all subsystems are larger than the correlation length and $\tilde{\xi} \approx \xi_\infty$, recovering bulk behavior. Close to criticality, where $\xi_\infty(\theta) \gg \langle L \rangle$, the correlation length in most subsystems is truncated at their boundaries and $\tilde{\xi}$ saturates toward $\langle L \rangle$, suppressing the divergence (Fig.~\ref{fig:theory}b). This saturation has a direct consequence for the freeze-out condition: at slow quench rates, the system approaches criticality more closely before freezing, so that $\xi_\infty$ enters the saturated regime and the frozen-in $\tilde{\xi}$ becomes only weakly dependent on $R$. The defect density $n \sim \tilde{\xi}(\theta_f)^{-d}$ therefore exhibits a reduced sensitivity to $R$ at slow quenches. At fast quench rates, freeze-out occurs far from criticality where $\xi_\infty \ll \langle L \rangle$, and the bulk KZ scaling $n \propto R^{\mu_{\mathrm{bulk}}}$ is recovered.

This transcendental equation, Eq.~(\ref{eq:freeze_out}), is solved numerically for each $R$, which yields the quenched defect density shown in Fig.~\ref{fig:theory}c (see Supplementary Sec.~\refSI{sec:parameters}{S2} for details of the parameterization). The resulting $n(R)$ curve displays a smooth, monotonic crossover from an effectively saturated regime ($\mu_{\mathrm{eff}} \rightarrow 0$) at the slowest quench rates to the bulk KZ scaling ($\mu_{\mathrm{eff}} \rightarrow \mu_{\mathrm{bulk}}$) at the fastest rates. Over any finite range of quench rates, this crossover may appear as an apparent power law with an intermediate effective exponent, $\mu_{\mathrm{eff}} < \mu_{\mathrm{bulk}}$, whose value is governed by the interplay between the freeze-out correlation length and the distribution of subsystem sizes.

The ensemble formulation naturally recovers the expected limiting cases. As $\langle L \rangle \rightarrow \infty$, the truncation becomes irrelevant and $\tilde{\xi} \rightarrow \xi_\infty$, recovering the conventional bulk KZ scaling with $\mu_{\mathrm{eff}}=\mu_{\mathrm{bulk}}$. In the opposite limit of a single bounded system, $f(L)=\delta(L-L_0)$, the sharp truncation from Eq.~(\ref{eq:xi_i}) provides an approximation to the smooth finite-size crossover described by rigorous scaling treatments~\cite{chandran2012kibble,liu2014dynamic}. Since the EFS-KZ framework averages over an ensemble of subsystem sizes, the microscopic details of the cutoff are statistically washed out, making the sharp-cutoff approximation well suited to heterogeneous systems such as polycrystals.

\section*{Polycrystals as test material}

\begin{figure}[!htbp]
\centering
\includegraphics[width=0.48\textwidth]{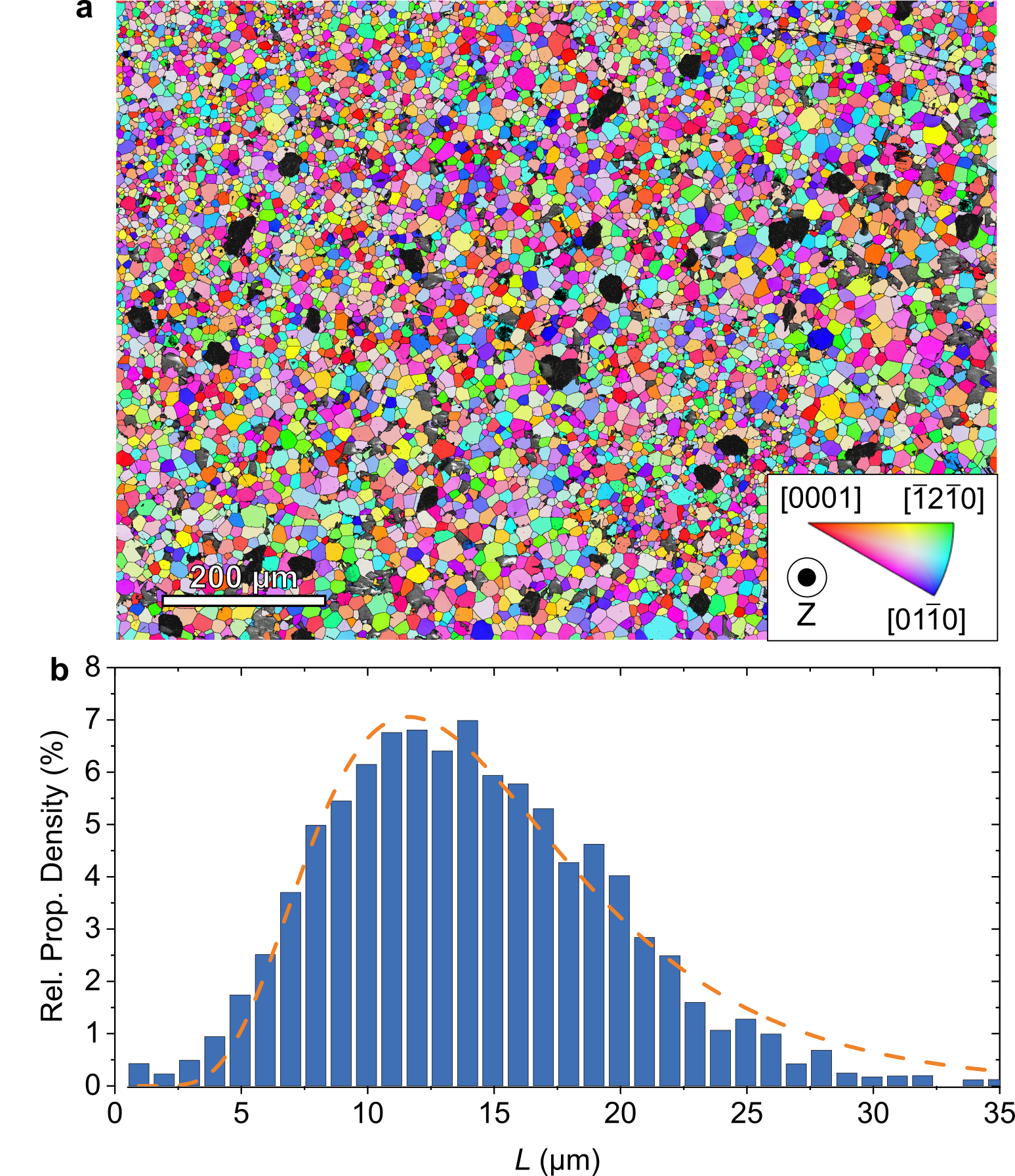}
\caption{
\textbf{Experimental realization of the ensemble-finite-size extension.} \textbf{(a)} EBSD map showing the crystallographic orientation of individual grains within the ErMnO$_3$ polycrystal. Black areas correspond to porosity of the sample. The data reveal a random $c$-axis orientation of the grains, effectively partitioning the microstructure into quasi-independent ferroelectric subsystems.  Grain orientations are expressed in Miller–Bravais notation $(hkil)$, appropriate for the $P6_{3}cm$ symmetry. \textbf{(b)} Relative  grain-size distribution extracted from the  dataset, illustrating the spread of grain sizes. A log-normal fit ((Eq.~\ref{eq:lognormal}), displayed as orange dashed line) describes the system within the EFS--KZ theoretical framework as $f(L)$, yielding a mean grain size of $\langle L \rangle = 14.9 \pm 0.2\,\mu\mathrm{m}$ and a standard deviation of $\sigma_L = 7.0 \pm 0.3\,\mu\mathrm{m}$.}
\label{fig:EBSD}
\end{figure}

To validate the EFS--KZ framework, we investigated the improper ferroelectric transition in polycrystalline hexagonal ErMnO$_3$, a canonical realization of a continuous symmetry-breaking transition with topologically protected vortex--antivortex defects~\cite{Choi2010,artyukhinLandauTheoryTopological2014b}. In this system, the trimerization transition at $T_c \approx 1430$~K breaks a continuous $U(1)$ symmetry, producing a vortex network that has previously been shown to follow KZ scaling in single crystals with exponent $\mu_{\mathrm{bulk}} \approx 0.58$, consistent with the 3D~XY universality class~\cite{Chae2012,Griffin2012,meier2017global}.

In contrast to single crystals, each grain in the polycrystal acts as a quasi-independent subsystem with finite characteristic size $L$. Experimentally, the trimerization order parameter vanishes at grain boundaries between misoriented grains, effectively imposing Dirichlet-like boundary conditions on each subsystem~\cite{schultheissConfinementDrivenInverseDomain2022,Hunnestad2023,Wolk2024}. These grains therefore constitute the statistical ensemble underlying the EFS--KZ framework and provide a natural platform for probing finite-size KZ dynamics. The sharp truncation introduced in Eq.~(\ref{eq:xi_i}) is physically motivated by this confinement. At the mean-field level, the Ginzburg--Pitaevskii analysis of the $\psi^4$ free energy under Dirichlet boundary conditions predicts a critical system size, $\overline{L}_s$, below which symmetry breaking is suppressed (Supplementary Sec.~\refSI{sec:TheoryExploreGP}{S3} and Figs.~\refSI{fig:critical_d}{S1} and~\refSI{fig:phi4_dirichlet}{S2}). Although this analysis does not capture the critical fluctuations of the 3D~XY universality class, the EFS--KZ framework averages over a broad distribution of subsystem sizes, rendering the predicted scaling insensitive to whether the crossover at an individual grain is perfectly sharp or smoothly rounded over a length scale small compared with the width of $f(L)$. Furthermore, the finite-size shift of the critical temperature is negligible ($\Delta T_c < 10^{-6}$\,K; Supplementary Sec.~\refSI{sec:DeltaTempFiniteSize}{S3 B}), so that all grains effectively undergo the transition at the same $T_c$.

\begin{figure*}[t]
\centering
\includegraphics[width=0.999\textwidth,height=0.50\textheight,keepaspectratio]{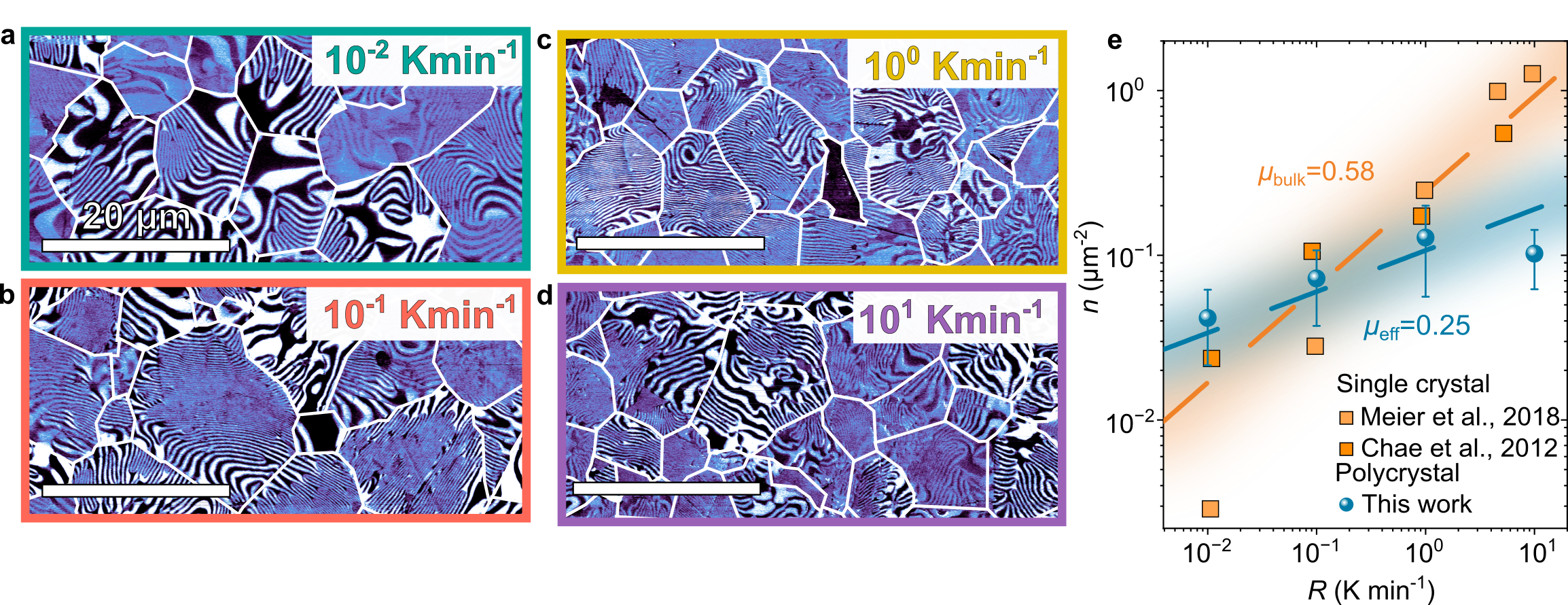}
  \vspace{-2mm}
\caption{
\textbf{Finite-size modification of KZ scaling in polycrystalline ErMnO$_3$.}
\textbf{(a–d)}~In-plane PFM images of the trimerization domain structure for polycrystalline samples cooled at $10^{-2}$ to $10^{1}$ \,K\,min$^{-1}$. White lines indicate grain boundaries, across which the trimerization order parameter vanishes, effectively partitioning the sample into quasi-independent confinement volumes. \textbf{(e)}~Vortex density, $n$, as a function of cooling rate, $R$, for the polycrystalline samples (blue circles), compared with single-crystal data (orange boxes) from Refs.~\cite{meier2017global,Chae2012}. The dashed blue curve shows the EFS–KZ cooling-rate dependence calculated using the experimentally measured grain-size distribution $f(L)$, displayed in Fig.~2b. Over the experimentally accessible cooling-rate window, the model predicts an apparent effective exponent consistent with the fit to the experimental vortex-density data ($\mu_{\mathrm{eff}} \approx 0.25$). The reduced cooling-rate dependence originates from finite-size truncation of the correlation length and statistical averaging over the grain-size distribution.}
\label{fig:combined}
\end{figure*}

We start by cooling polycrystals at different rates through $T_c$, as schematically visualized in Fig.~\refSI{fig:figS1}{S3}. The experiments are designed such that the crystal structure and overall microstructure remain unchanged across all cooling rates (Fig.~\refSI{fig:figS1a}{S4}), excluding structural variations as the origin of the observed scaling behaviour. To establish a quantitative link between the microstructure and our theoretical framework, grain-size distributions are extracted from an electron backscatter diffraction (EBSD) map (Fig.~\ref{fig:EBSD}a). The EBSD map shows a random orientation of grains, effectively decoupling neighboring grains through their crystallographic misorientation. The histogram (Fig.~\ref{fig:EBSD}b) is well described by a log-normal fit (Eq.~\ref{eq:lognormal}) and provides a direct quantification of $f(L)$ with $\langle L \rangle = 14.9 \pm 0.2\,\mu\mathrm{m}$ and a standard deviation $\sigma_L = 7.0 \pm 0.3\,\mu\mathrm{m}$. 
This experimentally determined distribution defines the ensemble of confinement lengths that limits the growth of the trimerization correlation length during the nonequilibrium transition. Using the measured grain-size distribution $f(L)$ to determine the finite-size crossover, the EFS--KZ framework predicts an apparent effective exponent of $\mu_{\mathrm{eff}}\approx0.25$ over the experimentally accessible cooling-rate window. This exponent is predicted by our EFS--KZ model without fitting to the experimental vortex-density data (Fig.~\ref{fig:combined}).

To test this prediction, we visualize the resulting vortex--antivortex network of the improper ferroelectric phase using piezoresponse force microscopy (PFM). As shown in Figs.~\ref{fig:combined}a--d, the domain morphology exhibits a pronounced cooling-rate dependence: slower cooling produces mixed vortex--stripe textures (Fig.~\ref{fig:combined}a), whereas faster quenches yield progressively finer vortex networks with reduced periodicity (Figs.~\ref{fig:combined}b--d). The coexistence of vortex-like and stripe-like domains is consistent with previous reports on polycrystalline hexagonal manganites~\cite{schultheissConfinementDrivenInverseDomain2022,Sandvik2023,Wolk2024}. Quantitatively, the stripe periodicity increases on average by a factor of two when the cooling rate is raised from $10^{-2}$ to $10^{0}$~K\,min$^{-1}$ (Fig.~\refSI{fig:figS2}{S5}), consistent with reduced domain annihilation and finite-size truncation at higher cooling rates. Note that the sample cooled at higher rates ($10^{1}$ \,K\,min$^{-1}$) is most likely influenced by anti-KZ behavior, as discussed in Ref.~\cite{meier2017global}.

Importantly, consistent with the random crystallographic orientations (Fig.~\ref{fig:EBSD}a), the trimerization order parameter vanishes at the grain boundaries, confirming that each grain acts as an effectively decoupled system during the ferroelectric phase transition~\cite{schultheissConfinementDrivenInverseDomain2022,Hunnestad2023,Wolk2024}. To quantify the KZ scaling, we determined the vortex density $n(R)$ from PFM images of samples quenched at different cooling rates (Figs.~\ref{fig:combined}a--d). Vortex cores were identified manually, normalized to the local grain area, and averaged over $\sim$20 grains per sample following the procedure of Refs.~\cite{Griffin2012,meier2017global}. As shown in Fig.~\ref{fig:combined}e, the experimentally determined vortex density increases by approximately 400\% as $R$ is raised from $10^{-2}$ to $10^{0}$~K\,min$^{-1}$. This cooling-rate dependence is substantially weaker than the bulk KZ scaling reported for single-crystal ErMnO$_3$ ($\mu_{\mathrm{bulk}}\approx0.58$)~\cite{Griffin2012,meier2017global} and closely follows the EFS--KZ prediction based on the measured grain-size distribution. In particular, the experimental $n(R)$ data are quantitatively consistent with our predicted effective exponent $\mu_{\mathrm{eff}}\approx0.25$ over the accessible cooling-rate window. Our model captures the reduced cooling-rate dependence of the measured vortex densities and yields an apparent exponent consistent with experiment by considering grain size distributions. The absolute density scale is set using the nonuniversal normalization parameter $\alpha=12$.


Our formulation adopts the minimal assumption that the correlation length is limited by geometry, \mbox{${\xi}_L \approx \min(\xi_\text{bulk}, L)$}, an approximation that is consistent with standard trends in finite-size scaling~\cite{Binder1987,fisher1972scaling}.  
Confinement replaces true singularities with smooth maxima, with Dirichlet boundary conditions imposed by grain interfaces.
Within this approximation, weakly coupled finite subsystems can be treated statistically without invoking a full renormalization-group treatment, consistent with finite-size analyses of nonequilibrium dynamics~\cite{chandran2012kibble,liu2014dynamic}.

Some caveats remain. Weak inter-grain coupling, anisotropic propagation of the order parameter, or local strain may modify quantitative prefactors. The influence of strain-induced anisotropic correlation lengths on vortex-density scaling is discussed in Supplementary Sec.~\refSI{sec:StrainAnisotropy}{S5} (Supplementary Fig.~\refSI{fig:figS3}{S6}), and thermalization effects were evaluated and found to be negligible compared to the imposed quench rates (Supplementary Sec.~\refSI{sec:ThermalizationEffects}{S6} and Fig.~\refSI{fig:figS4s}{S7}). Neither effect was found to alter the core result that statistical confinement systematically reduces the apparent exponent $\mu_{\mathrm{eff}}$ relative to $\mu_{\mathrm{bulk}}$.

Our findings highlight that universality in nonequilibrium dynamics is more flexible than its equilibrium counterpart: rather than being destroyed by confinement, it can acquire an ensemble-dependent statistical form governed by the distribution of finite subsystem sizes. This realization opens a path toward controlling nonequilibrium scaling through engineered microstructure, enabling quantitative tuning of topological-defect densities and relaxation dynamics across diverse material platforms. Beyond ferroelectrics, the ensemble-finite-size framework can be applied to confined superconductors, charge-density-wave systems, and ultracold gases where intrinsic geometric or disorder-induced cutoffs limit correlation growth.  
It thus provides a general foundation for understanding how confinement and heterogeneity reshape nonequilibrium universality in real materials.

\section{Methods}
\label{sec:Methods}

\subsection{Synthesis and sample preparation} 
\label{sec:Synthesis}
We synthesized phase-pure polycrystalline ErMnO$_3$ by solid-state reaction from stoichiometric \ce{Er2O3} and \ce{Mn2O3} precursors following established procedures (see Ref.~\cite{schultheissConfinementDrivenInverseDomain2022} for details). Pellets (10\,mm diameter, $\sim$1\,mm thickness) were sintered at $1750$\,K for 24\,h (heating rate 5\,K\,min$^{-1}$) to achieve densification (Entech Energiteknik AB, Ängelholm, Sweden). Controlled cooling through the improper ferroelectric transition at $T_c\!\approx\!1430$\,K~\cite{Chae2012} was performed with rates between $10^{-2}$ and $10^{1}$\,K\,min$^{-1}$ within $\pm$20\,K of $T_c$, while outside this interval a rate of 5\,K\,min$^{-1}$ was used (cooling profile shown in Fig.~\refSI{fig:figS1}{S3}). Phase purity was confirmed by X-ray diffraction (XRD, D8 ADVANCE, Bruker, Billerica, MA, USA). 
\subsection{Microstructural analysis} 
\label{sec:PFMmeasurements}
For local characterization of the microstructure the samples were lapped with a 9 µm-grained \ce{Al2O3} (Logitech Ltd, Glasgow, UK) water suspension. Subsequently, the samples were polished using a silica slurry (SF 1 Polishing Fluid, Logitech). 

EBSD data was collected using a Symmetry S3 detector (Oxford Instruments, High Wycombe, UK) on a
Hitachi SU3900SE thermally assisted FEG-SEM operated at 20\,kV with a probe current of $\sim 10$\,nA. 
Patterns were acquired at $\sim 500$\,Hz using AZtec version 6.3 (Oxford Instruments). The map is a montage of 57 individual fields, each covering $253 \times 190\,\mu\text{m}^2$.
EBSD patterns were acquired with a step size of $0.4\,\mu\mathrm{m}$. The resulting map (Fig.~2a) is cropped to make a rectangular map with an overall dimension of $0.79 \times 1.09\,\text{mm}^2$. The montaged data were improved using AZtecCrystal’s AutoClean with Void Detection (based on pattern quality). Grain detection used a critical misorientation angle of $10^\circ$, yielding 8894 grains. Only grains containing 10 pixels or more were included in the statistical analysis. 
Grains touching the border of the map were also excluded from the analysis to avoid skew towards smaller average grain size.

PFM measurements were performed on a NT-MDT Ntegra Prisma system (Spectrum Instruments Ltd., Limerick, Ireland) using an electrically conductive platinum tip (Spark 150 Pt, Nu Nano Ltd., Bristol, UK). The laser deflection was read out as the amplitude, $R$, and the phase, $\theta$, of the piezoresponse by two lock-in amplifiers (SR830, Stanford Research Systems, Sunnyvale, CA, USA). Spatial resolution of the domain structure is obtained by displaying $R \cos(\theta)$, to qualitatively distinguish domains with different orientation. Prior to the measurements, the PFM response was calibrated on a periodically poled out-of-plane \ce{LiNbO3} sample (PFM03, Spectrum Instruments Ltd., Limerick,
Ireland). PFM image analysis was done utilizing Gwyddion (v. 2.61, Brno, Czech Republic) and ImageJ (v. 1.53, Bethesda, MA, USA). The vortex density was calculated as $\rho = \frac{n}{A_{\mathrm{grain}}}$, where $n$ is the manually determined number of vortex cores for a grain with an area, $A_\mathrm{grain}$. To obtain a statistically relevant value of $\rho$, for each cooling rate about 20 grains were analyzed. The periodicity of stripe-like domains was calculated by averaging over 80 profiles for each cooling rate drawn through stripe-like domain structures. For each profile, the periodicity was identified automatically utilizing a fast Fourier transform.

\section{Acknowledgements}
\label{sec:Acknowledgements}
We acknowledge helpful discussions with G. Catalan. Theoretical work (G.M., I.C., S.G.) was funded by the U.S. Department of Energy, Office of Science, Office of Basic Energy Sciences, Materials Sciences and Engineering Division under Contract No.\ DE-AC02-05-CH11231 within the Theory of Materials program. Computational resources were provided by the National Energy Research Scientific Computing Center and the Molecular Foundry, DOE Office of Science User Facilities supported by the Office of Science, U.S.\ Department of Energy under Contract No.\ DEAC02-05CH11231. The work performed at the Molecular Foundry was supported by the Office of Science, Office of Basic Energy Sciences, of the U.S. Department of Energy under the same contract. G.M. acknowledges support from the Department of Energy Computational Science Graduate Fellowship (DOE CSGF) under grant DE-SC0020347. I.C. acknowledges a fellowship award under contract FA9550-21-F-0003 through the National Defense Science and Engineering Graduate (NDSEG) Fellowship Program. K.A.P. was supported by the U.S.\ Department of Energy, Office of Science,
Office of Basic Energy Sciences, Materials Sciences and Engineering Division, under Contract No. DE-AC02-05-C.H.11231 (Materials
Project Program No. KC23MP).  J.S. acknowledges the support of the Alexander von Humboldt Foundation through a Feodor-Lynen research fellowship, the German Academic Exchange Service (DAAD) for a Post-Doctoral Fellowship (Short-term program) and the Research Council of Norway through the Researcher Project for Scientific Renewal programme under Project No. 357912. D.M. thanks the NTNU for support through the Onsager Fellowship Program, the outstanding Academic Fellow Program, and D.M., J.H., and J.S. acknowledge funding from the European Research Council (ERC) under the European Union’s Horizon 2020 Research and Innovation Program (Grant Agreement No. 863691).

\section{Author contributions}
\label{sec:AuthorContribution}
J.S. and D.M. designed the project and experiments. M.H. performed the experiments under supervision of J.S. M.E.H. obtained and evaluated the EBSD data. J.H. provided supporting electron microscopy measurements, supervised by D.M. The theoretical work was performed by G.M. and I.C. under the supervision of S.G. J.S., I.C., G.M., D.M., and S.G. wrote the manuscript, with additions, restructuring, and revisions by coauthors. All authors provided feedback to the manuscript.

\section{Competing interests}
M.E.H. is an employee of Oxford Instruments, the manufacturer of the scanning electron microscope used for EBSD measurements reported in this study. The
remaining authors declare no competing interests.



\bibliography{references, references_orig}

\begin{thebibliography}{43}%
\makeatletter
\providecommand \@ifxundefined [1]{%
 \@ifx{#1\undefined}
}%
\providecommand \@ifnum [1]{%
 \ifnum #1\expandafter \@firstoftwo
 \else \expandafter \@secondoftwo
 \fi
}%
\providecommand \@ifx [1]{%
 \ifx #1\expandafter \@firstoftwo
 \else \expandafter \@secondoftwo
 \fi
}%
\providecommand \natexlab [1]{#1}%
\providecommand \enquote  [1]{``#1''}%
\providecommand \bibnamefont  [1]{#1}%
\providecommand \bibfnamefont [1]{#1}%
\providecommand \citenamefont [1]{#1}%
\providecommand \href@noop [0]{\@secondoftwo}%
\providecommand \href [0]{\begingroup \@sanitize@url \@href}%
\providecommand \@href[1]{\@@startlink{#1}\@@href}%
\providecommand \@@href[1]{\endgroup#1\@@endlink}%
\providecommand \@sanitize@url [0]{\catcode `\\12\catcode `\$12\catcode
  `\&12\catcode `\#12\catcode `\^12\catcode `\_12\catcode `\%12\relax}%
\providecommand \@@startlink[1]{}%
\providecommand \@@endlink[0]{}%
\providecommand \url  [0]{\begingroup\@sanitize@url \@url }%
\providecommand \@url [1]{\endgroup\@href {#1}{\urlprefix }}%
\providecommand \urlprefix  [0]{URL }%
\providecommand \Eprint [0]{\href }%
\providecommand \doibase [0]{https://doi.org/}%
\providecommand \selectlanguage [0]{\@gobble}%
\providecommand \bibinfo  [0]{\@secondoftwo}%
\providecommand \bibfield  [0]{\@secondoftwo}%
\providecommand \translation [1]{[#1]}%
\providecommand \BibitemOpen [0]{}%
\providecommand \bibitemStop [0]{}%
\providecommand \bibitemNoStop [0]{.\EOS\space}%
\providecommand \EOS [0]{\spacefactor3000\relax}%
\providecommand \BibitemShut  [1]{\csname bibitem#1\endcsname}%
\let\auto@bib@innerbib\@empty
\bibitem [{\citenamefont {Kibble}(1976)}]{Kibble1976}%
  \BibitemOpen
  \bibfield  {author} {\bibinfo {author} {\bibfnamefont {T.~W.~B.}\
  \bibnamefont {Kibble}},\ }\bibfield  {title} {\bibinfo {title} {Topology of
  cosmic domains and strings},\ }\href
  {https://doi.org/10.1088/0305-4470/9/8/029} {\bibfield  {journal} {\bibinfo
  {journal} {Journal of Physics A: Mathematical and General}\ }\textbf
  {\bibinfo {volume} {9}},\ \bibinfo {pages} {1387} (\bibinfo {year}
  {1976})}\BibitemShut {NoStop}%
\bibitem [{\citenamefont {Zurek}(1985)}]{zurek1985cosmological}%
  \BibitemOpen
  \bibfield  {author} {\bibinfo {author} {\bibfnamefont {W.~H.}\ \bibnamefont
  {Zurek}},\ }\bibfield  {title} {\bibinfo {title} {Cosmological experiments in
  superfluid helium?},\ }\href {https://doi.org/10.1038/317505a0} {\bibfield
  {journal} {\bibinfo  {journal} {Nature}\ }\textbf {\bibinfo {volume} {317}},\
  \bibinfo {pages} {505} (\bibinfo {year} {1985})}\BibitemShut {NoStop}%
\bibitem [{\citenamefont {Chuang}\ \emph {et~al.}(1991)\citenamefont {Chuang},
  \citenamefont {Durrer}, \citenamefont {Turok},\ and\ \citenamefont
  {Yurke}}]{Chuang1991}%
  \BibitemOpen
  \bibfield  {author} {\bibinfo {author} {\bibfnamefont {I.}~\bibnamefont
  {Chuang}}, \bibinfo {author} {\bibfnamefont {R.}~\bibnamefont {Durrer}},
  \bibinfo {author} {\bibfnamefont {N.}~\bibnamefont {Turok}},\ and\ \bibinfo
  {author} {\bibfnamefont {B.}~\bibnamefont {Yurke}},\ }\bibfield  {title}
  {\bibinfo {title} {Cosmology in the laboratory: Defect dynamics in liquid
  crystals},\ }\href {https://doi.org/10.1126/science.251.4999.1336} {\bibfield
   {journal} {\bibinfo  {journal} {Science}\ }\textbf {\bibinfo {volume}
  {251}},\ \bibinfo {pages} {1336} (\bibinfo {year} {1991})}\BibitemShut
  {NoStop}%
\bibitem [{\citenamefont {Ducci}\ \emph {et~al.}(1999)\citenamefont {Ducci},
  \citenamefont {Ramazza}, \citenamefont {Gonz{\'a}lez-Vi{\~n}as},\ and\
  \citenamefont {Arecchi}}]{Ducci1999}%
  \BibitemOpen
  \bibfield  {author} {\bibinfo {author} {\bibfnamefont {S.}~\bibnamefont
  {Ducci}}, \bibinfo {author} {\bibfnamefont {P.~L.}\ \bibnamefont {Ramazza}},
  \bibinfo {author} {\bibfnamefont {W.}~\bibnamefont
  {Gonz{\'a}lez-Vi{\~n}as}},\ and\ \bibinfo {author} {\bibfnamefont {F.~T.}\
  \bibnamefont {Arecchi}},\ }\bibfield  {title} {\bibinfo {title} {Order
  parameter fragmentation after a symmetry-breaking transition},\ }\href
  {https://doi.org/10.1103/PhysRevLett.83.5210} {\bibfield  {journal} {\bibinfo
   {journal} {Physical Review Letters}\ }\textbf {\bibinfo {volume} {83}},\
  \bibinfo {pages} {5210} (\bibinfo {year} {1999})}\BibitemShut {NoStop}%
\bibitem [{\citenamefont {Hendry}\ \emph {et~al.}(1994)\citenamefont {Hendry},
  \citenamefont {Lawson}, \citenamefont {Lee}, \citenamefont {McClintock},\
  and\ \citenamefont {Williams}}]{Hendry1994}%
  \BibitemOpen
  \bibfield  {author} {\bibinfo {author} {\bibfnamefont {P.~C.}\ \bibnamefont
  {Hendry}}, \bibinfo {author} {\bibfnamefont {N.~S.}\ \bibnamefont {Lawson}},
  \bibinfo {author} {\bibfnamefont {R.~A.~M.}\ \bibnamefont {Lee}}, \bibinfo
  {author} {\bibfnamefont {P.~V.~E.}\ \bibnamefont {McClintock}},\ and\
  \bibinfo {author} {\bibfnamefont {C.~D.~H.}\ \bibnamefont {Williams}},\
  }\bibfield  {title} {\bibinfo {title} {Generation of defects in superfluid
  {${}^4$He} as an analogue of the formation of cosmic strings},\ }\href
  {https://doi.org/10.1038/368315a0} {\bibfield  {journal} {\bibinfo  {journal}
  {Nature}\ }\textbf {\bibinfo {volume} {368}},\ \bibinfo {pages} {315}
  (\bibinfo {year} {1994})}\BibitemShut {NoStop}%
\bibitem [{\citenamefont {B{\"a}uerle}\ \emph {et~al.}(1996)\citenamefont
  {B{\"a}uerle}, \citenamefont {Bunkov}, \citenamefont {Fisher}, \citenamefont
  {Godfrin},\ and\ \citenamefont {Pickett}}]{Bauerle1996}%
  \BibitemOpen
  \bibfield  {author} {\bibinfo {author} {\bibfnamefont {C.}~\bibnamefont
  {B{\"a}uerle}}, \bibinfo {author} {\bibfnamefont {Y.~M.}\ \bibnamefont
  {Bunkov}}, \bibinfo {author} {\bibfnamefont {S.~N.}\ \bibnamefont {Fisher}},
  \bibinfo {author} {\bibfnamefont {H.}~\bibnamefont {Godfrin}},\ and\ \bibinfo
  {author} {\bibfnamefont {G.~R.}\ \bibnamefont {Pickett}},\ }\bibfield
  {title} {\bibinfo {title} {Laboratory simulation of cosmic string formation
  in the early {Universe} using superfluid {${}^3$He}},\ }\href
  {https://doi.org/10.1038/382332a0} {\bibfield  {journal} {\bibinfo  {journal}
  {Nature}\ }\textbf {\bibinfo {volume} {382}},\ \bibinfo {pages} {332}
  (\bibinfo {year} {1996})}\BibitemShut {NoStop}%
\bibitem [{\citenamefont {Carmi}\ \emph {et~al.}(2000)\citenamefont {Carmi},
  \citenamefont {Polturak},\ and\ \citenamefont {Koren}}]{Carmi2000}%
  \BibitemOpen
  \bibfield  {author} {\bibinfo {author} {\bibfnamefont {R.}~\bibnamefont
  {Carmi}}, \bibinfo {author} {\bibfnamefont {E.}~\bibnamefont {Polturak}},\
  and\ \bibinfo {author} {\bibfnamefont {G.}~\bibnamefont {Koren}},\ }\bibfield
   {title} {\bibinfo {title} {Observation of spontaneous flux generation in a
  multi-{Josephson}-junction loop},\ }\href
  {https://doi.org/10.1103/PhysRevLett.84.4966} {\bibfield  {journal} {\bibinfo
   {journal} {Physical Review Letters}\ }\textbf {\bibinfo {volume} {84}},\
  \bibinfo {pages} {4966} (\bibinfo {year} {2000})}\BibitemShut {NoStop}%
\bibitem [{\citenamefont {Ulm}\ \emph {et~al.}(2013)\citenamefont {Ulm},
  \citenamefont {Ro{\ss}nagel}, \citenamefont {Jacob}, \citenamefont
  {Deg{\"u}nther}, \citenamefont {Dawkins}, \citenamefont {Poschinger},
  \citenamefont {Nigmatullin}, \citenamefont {Retzker}, \citenamefont {Plenio},
  \citenamefont {Schmidt-Kaler},\ and\ \citenamefont {Singer}}]{Ulm2013}%
  \BibitemOpen
  \bibfield  {author} {\bibinfo {author} {\bibfnamefont {S.}~\bibnamefont
  {Ulm}}, \bibinfo {author} {\bibfnamefont {J.}~\bibnamefont {Ro{\ss}nagel}},
  \bibinfo {author} {\bibfnamefont {G.}~\bibnamefont {Jacob}}, \bibinfo
  {author} {\bibfnamefont {C.}~\bibnamefont {Deg{\"u}nther}}, \bibinfo {author}
  {\bibfnamefont {S.~T.}\ \bibnamefont {Dawkins}}, \bibinfo {author}
  {\bibfnamefont {U.~G.}\ \bibnamefont {Poschinger}}, \bibinfo {author}
  {\bibfnamefont {R.}~\bibnamefont {Nigmatullin}}, \bibinfo {author}
  {\bibfnamefont {A.}~\bibnamefont {Retzker}}, \bibinfo {author} {\bibfnamefont
  {M.~B.}\ \bibnamefont {Plenio}}, \bibinfo {author} {\bibfnamefont
  {F.}~\bibnamefont {Schmidt-Kaler}},\ and\ \bibinfo {author} {\bibfnamefont
  {K.}~\bibnamefont {Singer}},\ }\bibfield  {title} {\bibinfo {title}
  {Observation of the {Kibble}--{Zurek} scaling law for defect formation in ion
  crystals},\ }\href {https://doi.org/10.1038/ncomms3290} {\bibfield  {journal}
  {\bibinfo  {journal} {Nature Communications}\ }\textbf {\bibinfo {volume}
  {4}},\ \bibinfo {pages} {2290} (\bibinfo {year} {2013})}\BibitemShut
  {NoStop}%
\bibitem [{\citenamefont {Liu}\ \emph {et~al.}(2014)\citenamefont {Liu},
  \citenamefont {Polkovnikov},\ and\ \citenamefont {Sandvik}}]{liu2014dynamic}%
  \BibitemOpen
  \bibfield  {author} {\bibinfo {author} {\bibfnamefont {C.-W.}\ \bibnamefont
  {Liu}}, \bibinfo {author} {\bibfnamefont {A.}~\bibnamefont {Polkovnikov}},\
  and\ \bibinfo {author} {\bibfnamefont {A.~W.}\ \bibnamefont {Sandvik}},\
  }\bibfield  {title} {\bibinfo {title} {Dynamic scaling at classical phase
  transitions approached through nonequilibrium quenching},\ }\href
  {https://doi.org/10.1103/PhysRevB.89.054307} {\bibfield  {journal} {\bibinfo
  {journal} {Physical Review B}\ }\textbf {\bibinfo {volume} {89}},\ \bibinfo
  {pages} {054307} (\bibinfo {year} {2014})}\BibitemShut {NoStop}%
\bibitem [{\citenamefont {Chandran}\ \emph {et~al.}(2012)\citenamefont
  {Chandran}, \citenamefont {Erez}, \citenamefont {Gubser},\ and\ \citenamefont
  {Sondhi}}]{chandran2012kibble}%
  \BibitemOpen
  \bibfield  {author} {\bibinfo {author} {\bibfnamefont {A.}~\bibnamefont
  {Chandran}}, \bibinfo {author} {\bibfnamefont {A.}~\bibnamefont {Erez}},
  \bibinfo {author} {\bibfnamefont {S.~S.}\ \bibnamefont {Gubser}},\ and\
  \bibinfo {author} {\bibfnamefont {S.~L.}\ \bibnamefont {Sondhi}},\ }\bibfield
   {title} {\bibinfo {title} {{Kibble}--{Zurek} problem: Universality and the
  scaling limit},\ }\href {https://doi.org/10.1103/PhysRevB.86.064304}
  {\bibfield  {journal} {\bibinfo  {journal} {Physical Review B}\ }\textbf
  {\bibinfo {volume} {86}},\ \bibinfo {pages} {064304} (\bibinfo {year}
  {2012})}\BibitemShut {NoStop}%
\bibitem [{\citenamefont {Polkovnikov}(2005)}]{polkovnikov2005universal}%
  \BibitemOpen
  \bibfield  {author} {\bibinfo {author} {\bibfnamefont {A.}~\bibnamefont
  {Polkovnikov}},\ }\bibfield  {title} {\bibinfo {title} {Universal adiabatic
  dynamics in the vicinity of a quantum critical point},\ }\href
  {https://doi.org/10.1103/PhysRevB.72.161201} {\bibfield  {journal} {\bibinfo
  {journal} {Physical Review B}\ }\textbf {\bibinfo {volume} {72}},\ \bibinfo
  {pages} {174302} (\bibinfo {year} {2005})}\BibitemShut {NoStop}%
\bibitem [{\citenamefont {Deng}\ \emph {et~al.}(2008)\citenamefont {Deng},
  \citenamefont {Ortiz},\ and\ \citenamefont {Viola}}]{deng2009dynamical}%
  \BibitemOpen
  \bibfield  {author} {\bibinfo {author} {\bibfnamefont {S.}~\bibnamefont
  {Deng}}, \bibinfo {author} {\bibfnamefont {G.}~\bibnamefont {Ortiz}},\ and\
  \bibinfo {author} {\bibfnamefont {L.}~\bibnamefont {Viola}},\ }\bibfield
  {title} {\bibinfo {title} {Dynamical non-ergodic scaling in continuous
  finite-order quantum phase transitions},\ }\href
  {https://doi.org/10.1209/0295-5075/84/67008} {\bibfield  {journal} {\bibinfo
  {journal} {Europhysics Letters}\ }\textbf {\bibinfo {volume} {84}},\ \bibinfo
  {pages} {67008} (\bibinfo {year} {2008})}\BibitemShut {NoStop}%
\bibitem [{\citenamefont {De~Grandi}\ \emph {et~al.}(2011)\citenamefont
  {De~Grandi}, \citenamefont {Polkovnikov},\ and\ \citenamefont
  {Sandvik}}]{de2011universal}%
  \BibitemOpen
  \bibfield  {author} {\bibinfo {author} {\bibfnamefont {C.}~\bibnamefont
  {De~Grandi}}, \bibinfo {author} {\bibfnamefont {A.}~\bibnamefont
  {Polkovnikov}},\ and\ \bibinfo {author} {\bibfnamefont {A.~W.}\ \bibnamefont
  {Sandvik}},\ }\bibfield  {title} {\bibinfo {title} {Universal nonequilibrium
  quantum dynamics in imaginary time},\ }\href
  {https://doi.org/10.1103/PhysRevB.84.224303} {\bibfield  {journal} {\bibinfo
  {journal} {Physical Review B}\ }\textbf {\bibinfo {volume} {84}},\ \bibinfo
  {pages} {224303} (\bibinfo {year} {2011})}\BibitemShut {NoStop}%
\bibitem [{\citenamefont {Kolodrubetz}\ \emph {et~al.}(2012)\citenamefont
  {Kolodrubetz}, \citenamefont {Clark},\ and\ \citenamefont
  {Huse}}]{kolodrubetz2012nonequilibrium}%
  \BibitemOpen
  \bibfield  {author} {\bibinfo {author} {\bibfnamefont {M.}~\bibnamefont
  {Kolodrubetz}}, \bibinfo {author} {\bibfnamefont {B.~K.}\ \bibnamefont
  {Clark}},\ and\ \bibinfo {author} {\bibfnamefont {D.~A.}\ \bibnamefont
  {Huse}},\ }\bibfield  {title} {\bibinfo {title} {Nonequilibrium dynamic
  critical scaling of the quantum {Ising} chain},\ }\href
  {https://doi.org/10.1103/PhysRevLett.109.015701} {\bibfield  {journal}
  {\bibinfo  {journal} {Physical Review Letters}\ }\textbf {\bibinfo {volume}
  {109}},\ \bibinfo {pages} {015701} (\bibinfo {year} {2012})}\BibitemShut
  {NoStop}%
\bibitem [{\citenamefont {Liu}\ \emph {et~al.}(2013)\citenamefont {Liu},
  \citenamefont {Polkovnikov},\ and\ \citenamefont {Sandvik}}]{liu2013quasi}%
  \BibitemOpen
  \bibfield  {author} {\bibinfo {author} {\bibfnamefont {C.-W.}\ \bibnamefont
  {Liu}}, \bibinfo {author} {\bibfnamefont {A.}~\bibnamefont {Polkovnikov}},\
  and\ \bibinfo {author} {\bibfnamefont {A.~W.}\ \bibnamefont {Sandvik}},\
  }\bibfield  {title} {\bibinfo {title} {Quasi-adiabatic quantum {Monte}
  {Carlo} algorithm for quantum evolution in imaginary time},\ }\href
  {https://doi.org/10.1103/PhysRevB.87.174302} {\bibfield  {journal} {\bibinfo
  {journal} {Physical Review B}\ }\textbf {\bibinfo {volume} {87}},\ \bibinfo
  {pages} {174302} (\bibinfo {year} {2013})}\BibitemShut {NoStop}%
\bibitem [{\citenamefont {De~Grandi}\ \emph {et~al.}(2013)\citenamefont
  {De~Grandi}, \citenamefont {Polkovnikov},\ and\ \citenamefont
  {Sandvik}}]{de2013microscopic}%
  \BibitemOpen
  \bibfield  {author} {\bibinfo {author} {\bibfnamefont {C.}~\bibnamefont
  {De~Grandi}}, \bibinfo {author} {\bibfnamefont {A.}~\bibnamefont
  {Polkovnikov}},\ and\ \bibinfo {author} {\bibfnamefont {A.~W.}\ \bibnamefont
  {Sandvik}},\ }\bibfield  {title} {\bibinfo {title} {Microscopic theory of
  non-adiabatic response in real and imaginary time},\ }\href
  {https://doi.org/10.1088/0953-8984/25/40/404216} {\bibfield  {journal}
  {\bibinfo  {journal} {Journal of Physics: Condensed Matter}\ }\textbf
  {\bibinfo {volume} {25}},\ \bibinfo {pages} {404216} (\bibinfo {year}
  {2013})}\BibitemShut {NoStop}%
\bibitem [{\citenamefont {Zhong}\ and\ \citenamefont
  {Xu}(2005)}]{zhong2005dynamic}%
  \BibitemOpen
  \bibfield  {author} {\bibinfo {author} {\bibfnamefont {F.}~\bibnamefont
  {Zhong}}\ and\ \bibinfo {author} {\bibfnamefont {Z.}~\bibnamefont {Xu}},\
  }\bibfield  {title} {\bibinfo {title} {Dynamic {Monte} {Carlo}
  renormalization group determination of critical exponents with linearly
  changing temperature},\ }\href {https://doi.org/10.1103/PhysRevB.71.132402}
  {\bibfield  {journal} {\bibinfo  {journal} {Physical Review B}\ }\textbf
  {\bibinfo {volume} {71}},\ \bibinfo {pages} {132402} (\bibinfo {year}
  {2005})}\BibitemShut {NoStop}%
\bibitem [{\citenamefont {Gong}\ \emph {et~al.}(2010)\citenamefont {Gong},
  \citenamefont {Zhong}, \citenamefont {Huang},\ and\ \citenamefont
  {Fan}}]{gong2010finite}%
  \BibitemOpen
  \bibfield  {author} {\bibinfo {author} {\bibfnamefont {S.}~\bibnamefont
  {Gong}}, \bibinfo {author} {\bibfnamefont {F.}~\bibnamefont {Zhong}},
  \bibinfo {author} {\bibfnamefont {X.}~\bibnamefont {Huang}},\ and\ \bibinfo
  {author} {\bibfnamefont {S.}~\bibnamefont {Fan}},\ }\bibfield  {title}
  {\bibinfo {title} {Finite-time scaling via linear driving},\ }\href
  {https://doi.org/10.1088/1367-2630/12/4/043036} {\bibfield  {journal}
  {\bibinfo  {journal} {New Journal of Physics}\ }\textbf {\bibinfo {volume}
  {12}},\ \bibinfo {pages} {043036} (\bibinfo {year} {2010})}\BibitemShut
  {NoStop}%
\bibitem [{\citenamefont {Jeli{\'c}}\ and\ \citenamefont
  {Cugliandolo}(2011)}]{jelicQuenchDynamics2d2011}%
  \BibitemOpen
  \bibfield  {author} {\bibinfo {author} {\bibfnamefont {A.}~\bibnamefont
  {Jeli{\'c}}}\ and\ \bibinfo {author} {\bibfnamefont {L.~F.}\ \bibnamefont
  {Cugliandolo}},\ }\bibfield  {title} {\bibinfo {title} {Quench dynamics of
  the {2D} {XY} model},\ }\href
  {https://doi.org/10.1088/1742-5468/2011/02/P02032} {\bibfield  {journal}
  {\bibinfo  {journal} {Journal of Statistical Mechanics: Theory and
  Experiment}\ }\textbf {\bibinfo {volume} {2011}},\ \bibinfo {pages} {P02032}
  (\bibinfo {year} {2011})}\BibitemShut {NoStop}%
\bibitem [{\citenamefont {Rojas}\ and\ \citenamefont
  {Rutenberg}(1999)}]{rojasDynamicalScalingTwodimensional1999}%
  \BibitemOpen
  \bibfield  {author} {\bibinfo {author} {\bibfnamefont {F.}~\bibnamefont
  {Rojas}}\ and\ \bibinfo {author} {\bibfnamefont {A.~D.}\ \bibnamefont
  {Rutenberg}},\ }\bibfield  {title} {\bibinfo {title} {Dynamical scaling: The
  two-dimensional {XY} model following a quench},\ }\href
  {https://doi.org/10.1103/PhysRevE.60.212} {\bibfield  {journal} {\bibinfo
  {journal} {Physical Review E}\ }\textbf {\bibinfo {volume} {60}},\ \bibinfo
  {pages} {212} (\bibinfo {year} {1999})}\BibitemShut {NoStop}%
\bibitem [{\citenamefont {Yurke}\ \emph {et~al.}(1993)\citenamefont {Yurke},
  \citenamefont {Pargellis}, \citenamefont {Kovacs},\ and\ \citenamefont
  {Huse}}]{yurkeCoarseningDynamicsXY1993}%
  \BibitemOpen
  \bibfield  {author} {\bibinfo {author} {\bibfnamefont {B.}~\bibnamefont
  {Yurke}}, \bibinfo {author} {\bibfnamefont {A.~N.}\ \bibnamefont
  {Pargellis}}, \bibinfo {author} {\bibfnamefont {T.}~\bibnamefont {Kovacs}},\
  and\ \bibinfo {author} {\bibfnamefont {D.~A.}\ \bibnamefont {Huse}},\
  }\bibfield  {title} {\bibinfo {title} {Coarsening dynamics of the {XY}
  model},\ }\href {https://doi.org/10.1103/PhysRevE.47.1525} {\bibfield
  {journal} {\bibinfo  {journal} {Physical Review E}\ }\textbf {\bibinfo
  {volume} {47}},\ \bibinfo {pages} {1525} (\bibinfo {year}
  {1993})}\BibitemShut {NoStop}%
\bibitem [{\citenamefont {Bergmann}\ \emph {et~al.}(1998)\citenamefont
  {Bergmann}, \citenamefont {Shi}, \citenamefont {Queisser},\ and\
  \citenamefont {Krinke}}]{bergmannFormationPolycrystallineSilicon1998}%
  \BibitemOpen
  \bibfield  {author} {\bibinfo {author} {\bibfnamefont {R.~B.}\ \bibnamefont
  {Bergmann}}, \bibinfo {author} {\bibfnamefont {F.~G.}\ \bibnamefont {Shi}},
  \bibinfo {author} {\bibfnamefont {H.~J.}\ \bibnamefont {Queisser}},\ and\
  \bibinfo {author} {\bibfnamefont {J.}~\bibnamefont {Krinke}},\ }\bibfield
  {title} {\bibinfo {title} {Formation of polycrystalline silicon with
  log-normal grain size distribution},\ }\href
  {https://doi.org/10.1016/S0169-4332(97)00494-7} {\bibfield  {journal}
  {\bibinfo  {journal} {Applied Surface Science}\ }\textbf {\bibinfo {volume}
  {123--124}},\ \bibinfo {pages} {376} (\bibinfo {year} {1998})}\BibitemShut
  {NoStop}%
\bibitem [{\citenamefont {Griffin}\ \emph {et~al.}(2012)\citenamefont
  {Griffin}, \citenamefont {Lilienblum}, \citenamefont {Delaney}, \citenamefont
  {Kumagai}, \citenamefont {Fiebig},\ and\ \citenamefont
  {Spaldin}}]{Griffin2012}%
  \BibitemOpen
  \bibfield  {author} {\bibinfo {author} {\bibfnamefont {S.~M.}\ \bibnamefont
  {Griffin}}, \bibinfo {author} {\bibfnamefont {M.}~\bibnamefont {Lilienblum}},
  \bibinfo {author} {\bibfnamefont {K.~T.}\ \bibnamefont {Delaney}}, \bibinfo
  {author} {\bibfnamefont {Y.}~\bibnamefont {Kumagai}}, \bibinfo {author}
  {\bibfnamefont {M.}~\bibnamefont {Fiebig}},\ and\ \bibinfo {author}
  {\bibfnamefont {N.~A.}\ \bibnamefont {Spaldin}},\ }\bibfield  {title}
  {\bibinfo {title} {Scaling behavior and beyond equilibrium in the hexagonal
  manganites},\ }\href {https://doi.org/10.1103/PhysRevX.2.041022} {\bibfield
  {journal} {\bibinfo  {journal} {Physical Review X}\ }\textbf {\bibinfo
  {volume} {2}},\ \bibinfo {pages} {041022} (\bibinfo {year}
  {2012})}\BibitemShut {NoStop}%
\bibitem [{\citenamefont {Luscombe}\ \emph {et~al.}(1996)\citenamefont
  {Luscombe}, \citenamefont {Luban},\ and\ \citenamefont
  {Reynolds}}]{luscombeFinitesizeScalingGlauber1996}%
  \BibitemOpen
  \bibfield  {author} {\bibinfo {author} {\bibfnamefont {J.~H.}\ \bibnamefont
  {Luscombe}}, \bibinfo {author} {\bibfnamefont {M.}~\bibnamefont {Luban}},\
  and\ \bibinfo {author} {\bibfnamefont {J.~P.}\ \bibnamefont {Reynolds}},\
  }\bibfield  {title} {\bibinfo {title} {Finite-size scaling of the {Glauber}
  model of critical dynamics},\ }\href
  {https://doi.org/10.1103/PhysRevE.53.5852} {\bibfield  {journal} {\bibinfo
  {journal} {Physical Review E}\ }\textbf {\bibinfo {volume} {53}},\ \bibinfo
  {pages} {5852} (\bibinfo {year} {1996})}\BibitemShut {NoStop}%
\bibitem [{\citenamefont {Schulthei{\ss}}\ \emph {et~al.}(2022)\citenamefont
  {Schulthei{\ss}}, \citenamefont {Xue}, \citenamefont {Roede}, \citenamefont
  {{\AA}nes}, \citenamefont {Danmo}, \citenamefont {Selbach}, \citenamefont
  {Chen},\ and\ \citenamefont
  {Meier}}]{schultheissConfinementDrivenInverseDomain2022}%
  \BibitemOpen
  \bibfield  {author} {\bibinfo {author} {\bibfnamefont {J.}~\bibnamefont
  {Schulthei{\ss}}}, \bibinfo {author} {\bibfnamefont {F.}~\bibnamefont {Xue}},
  \bibinfo {author} {\bibfnamefont {E.}~\bibnamefont {Roede}}, \bibinfo
  {author} {\bibfnamefont {H.~W.}\ \bibnamefont {{\AA}nes}}, \bibinfo {author}
  {\bibfnamefont {F.~H.}\ \bibnamefont {Danmo}}, \bibinfo {author}
  {\bibfnamefont {S.~M.}\ \bibnamefont {Selbach}}, \bibinfo {author}
  {\bibfnamefont {L.-Q.}\ \bibnamefont {Chen}},\ and\ \bibinfo {author}
  {\bibfnamefont {D.}~\bibnamefont {Meier}},\ }\bibfield  {title} {\bibinfo
  {title} {Confinement-driven inverse domain scaling in polycrystalline
  {ErMnO$_3$}},\ }\href {https://doi.org/10.1002/adma.202203449} {\bibfield
  {journal} {\bibinfo  {journal} {Advanced Materials}\ }\textbf {\bibinfo
  {volume} {34}},\ \bibinfo {pages} {2203449} (\bibinfo {year}
  {2022})}\BibitemShut {NoStop}%
\bibitem [{\citenamefont {Hunnestad}\ \emph {et~al.}(2023)\citenamefont
  {Hunnestad}, \citenamefont {Schulthei{\ss}}, \citenamefont {Mathisen},
  \citenamefont {Ushakov}, \citenamefont {Hatzoglou}, \citenamefont {van
  Helvoort},\ and\ \citenamefont {Meier}}]{Hunnestad2023}%
  \BibitemOpen
  \bibfield  {author} {\bibinfo {author} {\bibfnamefont {K.~A.}\ \bibnamefont
  {Hunnestad}}, \bibinfo {author} {\bibfnamefont {J.}~\bibnamefont
  {Schulthei{\ss}}}, \bibinfo {author} {\bibfnamefont {A.~C.}\ \bibnamefont
  {Mathisen}}, \bibinfo {author} {\bibfnamefont {I.}~\bibnamefont {Ushakov}},
  \bibinfo {author} {\bibfnamefont {C.}~\bibnamefont {Hatzoglou}}, \bibinfo
  {author} {\bibfnamefont {A.~T.~J.}\ \bibnamefont {van Helvoort}},\ and\
  \bibinfo {author} {\bibfnamefont {D.}~\bibnamefont {Meier}},\ }\bibfield
  {title} {\bibinfo {title} {Quantitative mapping of chemical defects at
  charged grain boundaries in a ferroelectric oxide},\ }\href
  {https://doi.org/10.1002/adma.202302543} {\bibfield  {journal} {\bibinfo
  {journal} {Advanced Materials}\ }\textbf {\bibinfo {volume} {35}},\ \bibinfo
  {pages} {2302543} (\bibinfo {year} {2023})}\BibitemShut {NoStop}%
\bibitem [{\citenamefont {Wolk}\ \emph {et~al.}(2024)\citenamefont {Wolk},
  \citenamefont {Dragland}, \citenamefont {Chavez~Panduro}, \citenamefont
  {Hjelmstad}, \citenamefont {Richarz}, \citenamefont {Yan}, \citenamefont
  {Bourret}, \citenamefont {Hunnestad}, \citenamefont {Tzschaschel},
  \citenamefont {Schulthei{\ss}},\ and\ \citenamefont {Meier}}]{Wolk2024}%
  \BibitemOpen
  \bibfield  {author} {\bibinfo {author} {\bibfnamefont {K.}~\bibnamefont
  {Wolk}}, \bibinfo {author} {\bibfnamefont {R.~S.}\ \bibnamefont {Dragland}},
  \bibinfo {author} {\bibfnamefont {E.}~\bibnamefont {Chavez~Panduro}},
  \bibinfo {author} {\bibfnamefont {M.~E.}\ \bibnamefont {Hjelmstad}}, \bibinfo
  {author} {\bibfnamefont {L.}~\bibnamefont {Richarz}}, \bibinfo {author}
  {\bibfnamefont {Z.}~\bibnamefont {Yan}}, \bibinfo {author} {\bibfnamefont
  {E.}~\bibnamefont {Bourret}}, \bibinfo {author} {\bibfnamefont {K.~A.}\
  \bibnamefont {Hunnestad}}, \bibinfo {author} {\bibfnamefont {C.}~\bibnamefont
  {Tzschaschel}}, \bibinfo {author} {\bibfnamefont {J.}~\bibnamefont
  {Schulthei{\ss}}},\ and\ \bibinfo {author} {\bibfnamefont {D.}~\bibnamefont
  {Meier}},\ }\bibfield  {title} {\bibinfo {title} {Coexistence of multi-scale
  domains in ferroelectric polycrystals with non-uniform grain-size
  distributions},\ }\href {https://doi.org/10.1016/j.matt.2024.04.041}
  {\bibfield  {journal} {\bibinfo  {journal} {Matter}\ }\textbf {\bibinfo
  {volume} {7}},\ \bibinfo {pages} {3097} (\bibinfo {year} {2024})}\BibitemShut
  {NoStop}%
\bibitem [{\citenamefont {Choi}\ \emph {et~al.}(2010)\citenamefont {Choi},
  \citenamefont {Horibe}, \citenamefont {Yi}, \citenamefont {Choi},
  \citenamefont {Wu},\ and\ \citenamefont {Cheong}}]{Choi2010}%
  \BibitemOpen
  \bibfield  {author} {\bibinfo {author} {\bibfnamefont {T.}~\bibnamefont
  {Choi}}, \bibinfo {author} {\bibfnamefont {Y.}~\bibnamefont {Horibe}},
  \bibinfo {author} {\bibfnamefont {H.~T.}\ \bibnamefont {Yi}}, \bibinfo
  {author} {\bibfnamefont {Y.~J.}\ \bibnamefont {Choi}}, \bibinfo {author}
  {\bibfnamefont {W.}~\bibnamefont {Wu}},\ and\ \bibinfo {author}
  {\bibfnamefont {S.-W.}\ \bibnamefont {Cheong}},\ }\bibfield  {title}
  {\bibinfo {title} {Insulating interlocked ferroelectric and structural
  antiphase domain walls in multiferroic {YMnO$_3$}},\ }\href
  {https://doi.org/10.1038/nmat2632} {\bibfield  {journal} {\bibinfo  {journal}
  {Nature Materials}\ }\textbf {\bibinfo {volume} {9}},\ \bibinfo {pages} {253}
  (\bibinfo {year} {2010})}\BibitemShut {NoStop}%
\bibitem [{\citenamefont {Artyukhin}\ \emph {et~al.}(2014)\citenamefont
  {Artyukhin}, \citenamefont {Delaney}, \citenamefont {Spaldin},\ and\
  \citenamefont {Mostovoy}}]{artyukhinLandauTheoryTopological2014b}%
  \BibitemOpen
  \bibfield  {author} {\bibinfo {author} {\bibfnamefont {S.}~\bibnamefont
  {Artyukhin}}, \bibinfo {author} {\bibfnamefont {K.~T.}\ \bibnamefont
  {Delaney}}, \bibinfo {author} {\bibfnamefont {N.~A.}\ \bibnamefont
  {Spaldin}},\ and\ \bibinfo {author} {\bibfnamefont {M.}~\bibnamefont
  {Mostovoy}},\ }\bibfield  {title} {\bibinfo {title} {{Landau} theory of
  topological defects in multiferroic hexagonal manganites},\ }\href
  {https://doi.org/10.1038/nmat3786} {\bibfield  {journal} {\bibinfo  {journal}
  {Nature Materials}\ }\textbf {\bibinfo {volume} {13}},\ \bibinfo {pages} {42}
  (\bibinfo {year} {2014})}\BibitemShut {NoStop}%
\bibitem [{\citenamefont {Chae}\ \emph {et~al.}(2012)\citenamefont {Chae},
  \citenamefont {Lee}, \citenamefont {Horibe}, \citenamefont {Tanimura},
  \citenamefont {Mori}, \citenamefont {Gao}, \citenamefont {Carr},\ and\
  \citenamefont {Cheong}}]{Chae2012}%
  \BibitemOpen
  \bibfield  {author} {\bibinfo {author} {\bibfnamefont {S.~C.}\ \bibnamefont
  {Chae}}, \bibinfo {author} {\bibfnamefont {N.}~\bibnamefont {Lee}}, \bibinfo
  {author} {\bibfnamefont {Y.}~\bibnamefont {Horibe}}, \bibinfo {author}
  {\bibfnamefont {M.}~\bibnamefont {Tanimura}}, \bibinfo {author}
  {\bibfnamefont {S.}~\bibnamefont {Mori}}, \bibinfo {author} {\bibfnamefont
  {B.}~\bibnamefont {Gao}}, \bibinfo {author} {\bibfnamefont {S.}~\bibnamefont
  {Carr}},\ and\ \bibinfo {author} {\bibfnamefont {S.-W.}\ \bibnamefont
  {Cheong}},\ }\bibfield  {title} {\bibinfo {title} {Direct observation of the
  proliferation of ferroelectric loop domains and vortex-antivortex pairs},\
  }\href {https://doi.org/10.1103/PhysRevLett.108.167603} {\bibfield  {journal}
  {\bibinfo  {journal} {Physical Review Letters}\ }\textbf {\bibinfo {volume}
  {108}},\ \bibinfo {pages} {167603} (\bibinfo {year} {2012})}\BibitemShut
  {NoStop}%
\bibitem [{\citenamefont {Meier}\ \emph {et~al.}(2017)\citenamefont {Meier},
  \citenamefont {Lilienblum}, \citenamefont {Griffin}, \citenamefont {Conder},
  \citenamefont {Pomjakushina}, \citenamefont {Yan}, \citenamefont {Bourret},
  \citenamefont {Meier}, \citenamefont {Lichtenberg}, \citenamefont {Salje},
  \citenamefont {Spaldin}, \citenamefont {Fiebig},\ and\ \citenamefont
  {Cano}}]{meier2017global}%
  \BibitemOpen
  \bibfield  {author} {\bibinfo {author} {\bibfnamefont {Q.~N.}\ \bibnamefont
  {Meier}}, \bibinfo {author} {\bibfnamefont {M.}~\bibnamefont {Lilienblum}},
  \bibinfo {author} {\bibfnamefont {S.~M.}\ \bibnamefont {Griffin}}, \bibinfo
  {author} {\bibfnamefont {K.}~\bibnamefont {Conder}}, \bibinfo {author}
  {\bibfnamefont {E.}~\bibnamefont {Pomjakushina}}, \bibinfo {author}
  {\bibfnamefont {Z.}~\bibnamefont {Yan}}, \bibinfo {author} {\bibfnamefont
  {E.}~\bibnamefont {Bourret}}, \bibinfo {author} {\bibfnamefont
  {D.}~\bibnamefont {Meier}}, \bibinfo {author} {\bibfnamefont
  {F.}~\bibnamefont {Lichtenberg}}, \bibinfo {author} {\bibfnamefont
  {E.~K.~H.}\ \bibnamefont {Salje}}, \bibinfo {author} {\bibfnamefont {N.~A.}\
  \bibnamefont {Spaldin}}, \bibinfo {author} {\bibfnamefont {M.}~\bibnamefont
  {Fiebig}},\ and\ \bibinfo {author} {\bibfnamefont {A.}~\bibnamefont {Cano}},\
  }\bibfield  {title} {\bibinfo {title} {Global formation of topological
  defects in the multiferroic hexagonal manganites},\ }\href
  {https://doi.org/10.1103/PhysRevX.7.041014} {\bibfield  {journal} {\bibinfo
  {journal} {Physical Review X}\ }\textbf {\bibinfo {volume} {7}},\ \bibinfo
  {pages} {041014} (\bibinfo {year} {2017})}\BibitemShut {NoStop}%
\bibitem [{\citenamefont {Sandvik}\ \emph {et~al.}(2023)\citenamefont
  {Sandvik}, \citenamefont {M{\"u}ller}, \citenamefont {{\AA}nes},
  \citenamefont {Zahn}, \citenamefont {He}, \citenamefont {Fiebig},
  \citenamefont {Lottermoser}, \citenamefont {Rojac}, \citenamefont {Meier},\
  and\ \citenamefont {Schulthei{\ss}}}]{Sandvik2023}%
  \BibitemOpen
  \bibfield  {author} {\bibinfo {author} {\bibfnamefont {O.~W.}\ \bibnamefont
  {Sandvik}}, \bibinfo {author} {\bibfnamefont {A.~M.}\ \bibnamefont
  {M{\"u}ller}}, \bibinfo {author} {\bibfnamefont {H.~W.}\ \bibnamefont
  {{\AA}nes}}, \bibinfo {author} {\bibfnamefont {M.}~\bibnamefont {Zahn}},
  \bibinfo {author} {\bibfnamefont {J.}~\bibnamefont {He}}, \bibinfo {author}
  {\bibfnamefont {M.}~\bibnamefont {Fiebig}}, \bibinfo {author} {\bibfnamefont
  {T.}~\bibnamefont {Lottermoser}}, \bibinfo {author} {\bibfnamefont
  {T.}~\bibnamefont {Rojac}}, \bibinfo {author} {\bibfnamefont
  {D.}~\bibnamefont {Meier}},\ and\ \bibinfo {author} {\bibfnamefont
  {J.}~\bibnamefont {Schulthei{\ss}}},\ }\bibfield  {title} {\bibinfo {title}
  {Pressure-control of non-ferroelastic ferroelectric domains in {ErMnO$_3$}},\
  }\href {https://doi.org/10.1021/acs.nanolett.3c01638} {\bibfield  {journal}
  {\bibinfo  {journal} {Nano Letters}\ }\textbf {\bibinfo {volume} {23}},\
  \bibinfo {pages} {6994} (\bibinfo {year} {2023})}\BibitemShut {NoStop}%
\bibitem [{\citenamefont {Binder}(1987)}]{Binder1987}%
  \BibitemOpen
  \bibfield  {author} {\bibinfo {author} {\bibfnamefont {K.}~\bibnamefont
  {Binder}},\ }\bibfield  {title} {\bibinfo {title} {Finite size effects on
  phase transitions},\ }\href {https://doi.org/10.1080/00150198708227908}
  {\bibfield  {journal} {\bibinfo  {journal} {Ferroelectrics}\ }\textbf
  {\bibinfo {volume} {73}},\ \bibinfo {pages} {43} (\bibinfo {year}
  {1987})}\BibitemShut {NoStop}%
\bibitem [{\citenamefont {Fisher}\ and\ \citenamefont
  {Barber}(1972)}]{fisher1972scaling}%
  \BibitemOpen
  \bibfield  {author} {\bibinfo {author} {\bibfnamefont {M.~E.}\ \bibnamefont
  {Fisher}}\ and\ \bibinfo {author} {\bibfnamefont {M.~N.}\ \bibnamefont
  {Barber}},\ }\bibfield  {title} {\bibinfo {title} {Scaling theory for
  finite-size effects in the critical region},\ }\href
  {https://doi.org/10.1103/PhysRevLett.28.1516} {\bibfield  {journal} {\bibinfo
   {journal} {Physical Review Letters}\ }\textbf {\bibinfo {volume} {28}},\
  \bibinfo {pages} {1516} (\bibinfo {year} {1972})}\BibitemShut {NoStop}%
\bibitem [{\citenamefont {Virtanen}\ \emph {et~al.}(2020)\citenamefont
  {Virtanen}, \citenamefont {Gommers}, \citenamefont {Oliphant}, \citenamefont
  {Haberland}, \citenamefont {Reddy}, \citenamefont {Cournapeau}, \citenamefont
  {Burovski}, \citenamefont {Peterson}, \citenamefont {Weckesser},
  \citenamefont {Bright} \emph {et~al.}}]{virtanen2020scipy}%
  \BibitemOpen
  \bibfield  {author} {\bibinfo {author} {\bibfnamefont {P.}~\bibnamefont
  {Virtanen}}, \bibinfo {author} {\bibfnamefont {R.}~\bibnamefont {Gommers}},
  \bibinfo {author} {\bibfnamefont {T.~E.}\ \bibnamefont {Oliphant}}, \bibinfo
  {author} {\bibfnamefont {M.}~\bibnamefont {Haberland}}, \bibinfo {author}
  {\bibfnamefont {T.}~\bibnamefont {Reddy}}, \bibinfo {author} {\bibfnamefont
  {D.}~\bibnamefont {Cournapeau}}, \bibinfo {author} {\bibfnamefont
  {E.}~\bibnamefont {Burovski}}, \bibinfo {author} {\bibfnamefont
  {P.}~\bibnamefont {Peterson}}, \bibinfo {author} {\bibfnamefont
  {W.}~\bibnamefont {Weckesser}}, \bibinfo {author} {\bibfnamefont
  {J.}~\bibnamefont {Bright}}, \emph {et~al.},\ }\bibfield  {title} {\bibinfo
  {title} {{SciPy} 1.0: fundamental algorithms for scientific computing in
  {Python}},\ }\href {https://doi.org/10.1038/s41592-019-0686-2} {\bibfield
  {journal} {\bibinfo  {journal} {Nature Methods}\ }\textbf {\bibinfo {volume}
  {17}},\ \bibinfo {pages} {261} (\bibinfo {year} {2020})}\BibitemShut
  {NoStop}%
\bibitem [{\citenamefont {Ginzburg}\ and\ \citenamefont
  {Pitaevskii}(1958)}]{ginzburg1958theory}%
  \BibitemOpen
  \bibfield  {author} {\bibinfo {author} {\bibfnamefont {V.~L.}\ \bibnamefont
  {Ginzburg}}\ and\ \bibinfo {author} {\bibfnamefont {L.~P.}\ \bibnamefont
  {Pitaevskii}},\ }\bibfield  {title} {\bibinfo {title} {On the theory of
  superfluidity},\ }\href@noop {} {\bibfield  {journal} {\bibinfo  {journal}
  {Soviet Physics JETP}\ }\textbf {\bibinfo {volume} {7}},\ \bibinfo {pages}
  {858} (\bibinfo {year} {1958})}\BibitemShut {NoStop}%
\bibitem [{\citenamefont {Kardar}(2007)}]{kardarStatisticalPhysicsFields2007}%
  \BibitemOpen
  \bibfield  {author} {\bibinfo {author} {\bibfnamefont {M.}~\bibnamefont
  {Kardar}},\ }\href {https://doi.org/10.1017/CBO9780511815881} {\emph
  {\bibinfo {title} {Statistical Physics of Fields}}}\ (\bibinfo  {publisher}
  {Cambridge University Press},\ \bibinfo {address} {Cambridge},\ \bibinfo
  {year} {2007})\BibitemShut {NoStop}%
\bibitem [{\citenamefont {Tomczyk}\ \emph {et~al.}(2012)\citenamefont
  {Tomczyk}, \citenamefont {Senos}, \citenamefont {Vilarinho},\ and\
  \citenamefont {Reaney}}]{Tomczyk2012}%
  \BibitemOpen
  \bibfield  {author} {\bibinfo {author} {\bibfnamefont {M.}~\bibnamefont
  {Tomczyk}}, \bibinfo {author} {\bibfnamefont {A.~M.}\ \bibnamefont {Senos}},
  \bibinfo {author} {\bibfnamefont {P.~M.}\ \bibnamefont {Vilarinho}},\ and\
  \bibinfo {author} {\bibfnamefont {I.~M.}\ \bibnamefont {Reaney}},\ }\bibfield
   {title} {\bibinfo {title} {Origin of microcracking in {YMnO$_3$} ceramics},\
  }\href {https://doi.org/10.1016/j.scriptamat.2011.11.014} {\bibfield
  {journal} {\bibinfo  {journal} {Scripta Materialia}\ }\textbf {\bibinfo
  {volume} {66}},\ \bibinfo {pages} {288} (\bibinfo {year} {2012})}\BibitemShut
  {NoStop}%
\bibitem [{\citenamefont {Xue}\ \emph {et~al.}(2017)\citenamefont {Xue},
  \citenamefont {Wang}, \citenamefont {Shi}, \citenamefont {Cheong},\ and\
  \citenamefont {Chen}}]{Xue2017}%
  \BibitemOpen
  \bibfield  {author} {\bibinfo {author} {\bibfnamefont {F.}~\bibnamefont
  {Xue}}, \bibinfo {author} {\bibfnamefont {X.~Y.}\ \bibnamefont {Wang}},
  \bibinfo {author} {\bibfnamefont {Y.}~\bibnamefont {Shi}}, \bibinfo {author}
  {\bibfnamefont {S.-W.}\ \bibnamefont {Cheong}},\ and\ \bibinfo {author}
  {\bibfnamefont {L.-Q.}\ \bibnamefont {Chen}},\ }\bibfield  {title} {\bibinfo
  {title} {Strain-induced incommensurate phases in hexagonal manganites},\
  }\href {https://doi.org/10.1103/PhysRevB.96.104109} {\bibfield  {journal}
  {\bibinfo  {journal} {Physical Review B}\ }\textbf {\bibinfo {volume} {96}},\
  \bibinfo {pages} {104109} (\bibinfo {year} {2017})}\BibitemShut {NoStop}%
\bibitem [{\citenamefont {Nalbach}\ \emph {et~al.}(2015)\citenamefont
  {Nalbach}, \citenamefont {Vishveshwara},\ and\ \citenamefont
  {Clerk}}]{nalbach2015quantum}%
  \BibitemOpen
  \bibfield  {author} {\bibinfo {author} {\bibfnamefont {P.}~\bibnamefont
  {Nalbach}}, \bibinfo {author} {\bibfnamefont {S.}~\bibnamefont
  {Vishveshwara}},\ and\ \bibinfo {author} {\bibfnamefont {A.~A.}\ \bibnamefont
  {Clerk}},\ }\bibfield  {title} {\bibinfo {title} {Quantum {Kibble}--{Zurek}
  physics in the presence of spatially correlated dissipation},\ }\href
  {https://doi.org/10.1103/PhysRevB.92.014306} {\bibfield  {journal} {\bibinfo
  {journal} {Physical Review B}\ }\textbf {\bibinfo {volume} {92}},\ \bibinfo
  {pages} {014306} (\bibinfo {year} {2015})}\BibitemShut {NoStop}%
\bibitem [{\citenamefont {Pang}\ \emph {et~al.}(2023)\citenamefont {Pang},
  \citenamefont {Li}, \citenamefont {Gao}, \citenamefont {Qian}, \citenamefont
  {Wang}, \citenamefont {Hong},\ and\ \citenamefont {Jiang}}]{pang2023thermal}%
  \BibitemOpen
  \bibfield  {author} {\bibinfo {author} {\bibfnamefont {Y.}~\bibnamefont
  {Pang}}, \bibinfo {author} {\bibfnamefont {Y.}~\bibnamefont {Li}}, \bibinfo
  {author} {\bibfnamefont {Z.}~\bibnamefont {Gao}}, \bibinfo {author}
  {\bibfnamefont {X.}~\bibnamefont {Qian}}, \bibinfo {author} {\bibfnamefont
  {X.}~\bibnamefont {Wang}}, \bibinfo {author} {\bibfnamefont {J.}~\bibnamefont
  {Hong}},\ and\ \bibinfo {author} {\bibfnamefont {P.}~\bibnamefont {Jiang}},\
  }\bibfield  {title} {\bibinfo {title} {Thermal transport manipulated by
  vortex domain walls in bulk {$h$-ErMnO$_3$}},\ }\href
  {https://doi.org/10.1016/j.mtphys.2023.100972} {\bibfield  {journal}
  {\bibinfo  {journal} {Materials Today Physics}\ }\textbf {\bibinfo {volume}
  {31}},\ \bibinfo {pages} {100972} (\bibinfo {year} {2023})}\BibitemShut
  {NoStop}%
\bibitem [{\citenamefont {Satoh}\ \emph {et~al.}(1995)\citenamefont {Satoh},
  \citenamefont {Shoji}, \citenamefont {Iwasaki},\ and\ \citenamefont
  {Kamegashira}}]{satoh1995high}%
  \BibitemOpen
  \bibfield  {author} {\bibinfo {author} {\bibfnamefont {H.}~\bibnamefont
  {Satoh}}, \bibinfo {author} {\bibfnamefont {T.}~\bibnamefont {Shoji}},
  \bibinfo {author} {\bibfnamefont {J.}~\bibnamefont {Iwasaki}},\ and\ \bibinfo
  {author} {\bibfnamefont {N.}~\bibnamefont {Kamegashira}},\ }\bibfield
  {title} {\bibinfo {title} {High-temperature enthalpy and heat capacity of
  {ErMnO$_3$}},\ }\href {https://doi.org/10.1016/0040-6031(95)02371-8}
  {\bibfield  {journal} {\bibinfo  {journal} {Thermochimica Acta}\ }\textbf
  {\bibinfo {volume} {261}},\ \bibinfo {pages} {47} (\bibinfo {year}
  {1995})}\BibitemShut {NoStop}%
\bibitem [{\citenamefont {Belrhiti-Nejjar}\ \emph {et~al.}(2025)\citenamefont
  {Belrhiti-Nejjar}, \citenamefont {Zahn}, \citenamefont {Limelette},
  \citenamefont {Haas}, \citenamefont {F{\'e}ger}, \citenamefont
  {Monot-Laffez}, \citenamefont {Horny}, \citenamefont {Meier}, \citenamefont
  {Giovannelli}, \citenamefont {Schulthei{\ss}},\ and\ \citenamefont
  {Nataf}}]{belrhiti2025domain}%
  \BibitemOpen
  \bibfield  {author} {\bibinfo {author} {\bibfnamefont {R.}~\bibnamefont
  {Belrhiti-Nejjar}}, \bibinfo {author} {\bibfnamefont {M.}~\bibnamefont
  {Zahn}}, \bibinfo {author} {\bibfnamefont {P.}~\bibnamefont {Limelette}},
  \bibinfo {author} {\bibfnamefont {M.}~\bibnamefont {Haas}}, \bibinfo {author}
  {\bibfnamefont {L.}~\bibnamefont {F{\'e}ger}}, \bibinfo {author}
  {\bibfnamefont {I.}~\bibnamefont {Monot-Laffez}}, \bibinfo {author}
  {\bibfnamefont {N.}~\bibnamefont {Horny}}, \bibinfo {author} {\bibfnamefont
  {D.}~\bibnamefont {Meier}}, \bibinfo {author} {\bibfnamefont
  {F.}~\bibnamefont {Giovannelli}}, \bibinfo {author} {\bibfnamefont
  {J.}~\bibnamefont {Schulthei{\ss}}},\ and\ \bibinfo {author} {\bibfnamefont
  {G.~F.}\ \bibnamefont {Nataf}},\ }\bibfield  {title} {\bibinfo {title}
  {Domain-wall driven suppression of thermal conductivity in a ferroelectric
  polycrystal},\ }\href {https://doi.org/10.1002/advs.202506931} {\bibfield
  {journal} {\bibinfo  {journal} {Advanced Science}\ }\textbf {\bibinfo
  {volume} {12}},\ \bibinfo {pages} {e06931} (\bibinfo {year}
  {2025})}\BibitemShut {NoStop}%
\end{thebibliography}%


\ifx\SubmitToggle\undefined
\else
  \if\SubmitToggle1
  \renewcommand{\thesection}{S\arabic{section}}
\renewcommand{\thesubsection}{\Alph{subsection}}
\renewcommand{\theequation}{S\arabic{equation}}
\renewcommand{\thetable}{S\arabic{table}}
\renewcommand{\thefigure}{S\arabic{figure}}

\setcounter{section}{0}
\setcounter{equation}{0}
\setcounter{table}{0}
\setcounter{figure}{0}

\begin{widetext}

\clearpage


\begin{center}

\vspace{4ex}

{
\LARGE
Supplementary information: \\ 

Kibble–Zurek dynamics beyond the thermodynamic limit \\

}

\vspace{4ex}
\vspace{4ex}

{
\large
Max Haas$^*$, 
Guy C. Moore$^*$, \\
Isaac M. Craig, 
Kristin A. Persson, 
Jiali He,
Michael E. Hjelmstad, \\
Dennis Meier, 
Sinéad M. Griffin,
Jan Schultheiß \\
}

\vspace{4ex}

\today

\vspace{4ex}

\end{center}


\section{Kibble-Zurek scaling over an ensemble of finite systems}
\label{sec:KZfinite}

We introduce a simple statistical model that describes the influence of finite size effects on Kibble Zurek (KZ) scaling over a disordered ensemble of finite systems. 


\subsection{Kibble-Zurek scaling}
\label{sec:kz_subtle}


Following from Ref.~\cite{Griffin2012}, the linear temperature quench is described as $T = T_0 - R \times t$, where $T_0$ is the initial temperature at time $t=0$, and $R$ is the quench rate. Solving for $t$, we arrive at $t = \frac{1}{R} \left( T_0 - T \right)$.
If $T_0 = T_C$, $\theta = 1 - T/T_C$.  The previously described conventional approach is to determine the frozen-in temperature, $T_f$, as the intersection point between the quench time curve, $t$, and relaxation time, $\tau$,
\begin{align}
    \tau(\theta_f) &= t_f = \frac{T_C}{R} \theta_f.
    \label{eq:tau_froz_temp}
\end{align}
However, more rigorously, following from Ref.~\cite{Griffin2012} one can determine $T_f$ using the equality between the correlation length, and the characteristic speed, $c(T) = \xi(T) / \tau(T)$. In this case, the transcendental equation involves an integral over time 
\begin{equation}
\begin{aligned}
    \xi(\theta_f) &= L_{\text{prop}}(\theta_f) \\
    &= \int_0^t dt' \ c[T(t')] 
    = \frac{T_C}{R} \int_0^{\theta_f} d\theta \ c(\theta) \\
    &= \frac{T_C}{R} \int_0^{\theta_f} d\theta \ \frac{\xi(\theta)}{\tau(\theta)},
\end{aligned}
\label{eq:xi_speed_froz_temp}
\end{equation}
where $L_{\text{prop}}$ is the information propagation length, i.e., the integrated effective speed of sound in the system \cite{Griffin2012}. It can be shown that for power law critical scaling of the correlation length and relaxation times,
\begin{align}
    \xi(\theta) &= \xi_0 \theta^{-\nu} \\
    \tau(\theta) &= \tau_0 \theta^{-z\nu}
\end{align}
that both Equations \ref{eq:tau_froz_temp} and \ref{eq:xi_speed_froz_temp} yield an identical KZ power law exponent, with the notable exception that each has a different power law coefficient \cite{Griffin2012}. This formulation additionally allows for the use of more general $\xi(\theta)$ and $\tau(\theta)$ expressions such as those shown in the following section. 

\subsection{Finite-size ensemble}
\label{sec:math_details}

If $f(L)$ is the probability distribution function of effective grain sizes, $L$, then one can consider the correlation length of a polycrystalline system, approximated as the expectation over grain size distributions
\begin{align}
    \tilde \xi (\theta) = \int_0^\infty dL \ f(L) \xi_L (\theta).
\end{align}
We choose $f(L)$ to be log-normal, in accordance with other studies, e.g. Ref.~\cite{bergmannFormationPolycrystallineSilicon1998}, assuming that random grain growth dominates the sintering process. $\xi_L(\theta)$ is the correlation length of an individual grain, truncated at the grain size $L$,
%
\begin{align}
    \xi_L (\theta)=
    \begin{cases}
    \xi_\infty(\theta) & L > \xi_\infty(\theta) \\
    L & L \leq \xi_\infty(\theta).
    \end{cases}
\end{align}
%


As discussed previously, the exact form of $\xi_L (\theta)$ in between both asymptotes ($\xi << L$ and $\xi \sim L$) will be averaged over. Therefore we have simply chosen $\xi_L (\theta)$ to be a sharp truncation of the correlation length, for any values of the correlation length that exceed the system size $L$. While we do not claim this is generally true, we do however note that there is a critical system size below which any symmetry-breaking Goldstone modes are damped such that an abrupt change in the correlation length could be justified (see Section~\ref{sec:GPsuperfluid}). Additionally, abrupt changes in critical scaling have been seen in Monte Carlo simulations under some - but not all - update dynamics \cite{liu2014dynamic}. Evaluating the integral above results in the following expression
\begin{equation}
\begin{aligned}
    \tilde \xi (\theta) &= \int_0^{\xi_\infty (\theta)} dL \ L f(L) + \int_{\xi_\infty(\theta)}^\infty dL \ \xi_\infty(\theta) f(L) \\ 
    &= \overline r[\xi_\infty(\theta)] + \xi_\infty(\theta)\left\{1 - F[\xi_\infty(\theta)]\right\},
\end{aligned}
\label{eq:eff_corr_length}
\end{equation}
where $F(L)$ is the cumulative distribution function (CDF) for the distribution of grain sizes. Within an effective-medium description of the ensemble, we assume that the relaxation time is related to the effective correlation length via the dynamical exponent $z$,
\begin{align}
    \tilde \tau(\theta) = \tilde \xi(\theta)^z.
\end{align}
We therefore assume that the dynamical scaling form is left unaffected by the length truncation, which agrees with previous theoretical work \cite{luscombeFinitesizeScalingGlauber1996}. 
The associated Kibble-Zurek scaling may then be obtained by using $\tilde \xi (\theta)$ and $\tilde \tau (\theta)$ in place of $\xi(\theta)$ and $\tau(\theta)$ in Equation~\ref{eq:xi_speed_froz_temp}. The resulting transcendental equation is solved numerically (see Sec.~\ref{sec:parameters} for further details).

\subsection{Dimensionless ensemble formulation}
\label{sec:dimensionless_formulation}

Casting the self-consistent freeze-out condition into dimensionless form,
\begin{align}
\bar{\xi}\left(\theta_f\right)=\bar{\xi}_f 
= \bar{R}^{-1} \int_0^{\theta_f} d \theta \ \frac{\bar{\xi}(\theta)}{\bar{\tau}(\theta)} \label{eq:n_efs}
\end{align}
where 
\begin{align}
\bar{\tau}(\theta) &= \tau(\theta) / \tau_0, \\
\bar{R} &= \left(\tau_0 / T_c\right) R,
\end{align}
and defining normalized subsystem size $x = L / \bar{L}$ and bare dimensionless correlation length $\bar{\xi}_0 = \xi_0 / \bar{L}$,
\begin{align}
\bar{\xi}(\theta) &= \int_0^{\infty} dx \ f(x) \bar{\xi}_L(\theta), \\
\bar{\xi}_L(\theta) &= \min \left(x, \bar{\xi}_{\infty}(\theta)\right), \\
\bar{\xi}_{\infty}(\theta) &= \bar{\xi}_0 \theta^{-\nu},
\end{align}
with the probability distribution function normalized as
\begin{align}
1 &= \int_0^{\infty} dx \ f(x), \\
1 &= \int_0^{\infty} dx \ x f(x).
\end{align}
In these normalized units, we set $\bar{\xi}_0 = 1$ and $\tau_0 / T_c = 1$, so that the shape of the crossover is governed solely by the critical exponents $(\nu, z)$ and the normalized subsystem-size distribution $f(x)$.

\section{Parameterization and Numerical details}
\label{sec:parameters}

To evaluate the predictions of the EFS--KZ model numerically, we solve the dimensionless self-consistent freeze-out integral using SciPy numerical integration and root-finding routines \cite{virtanen2020scipy}. The physical and empirical parameters used in the calculations and comparison to experiment are summarized below:

\begin{itemize}
    \item \textbf{Phase transition temperature}: $T_C = 1430 \ \text{K}$ for \ce{ErMnO3}.
    \item \textbf{3D XY universality exponents}: Correlation length exponent $\nu = 0.672$ and dynamical critical exponent $z = 2.0$, yielding the bulk single-crystal Kibble--Zurek exponent $\mu_{\mathrm{bulk}} = \frac{\nu d}{1 + z \nu} = 0.58$ in $d = 2$ spatial dimensions \cite{meier2017global, Griffin2012}.
    \item \textbf{EBSD grain-size distribution}: The probability distribution function $f(L)$ is fitted to EBSD map measurements on polycrystalline \ce{ErMnO3}, yielding an arithmetic mean grain diameter $\langle L \rangle = 14.9 \pm 0.2\,\mu\mathrm{m}$ and standard deviation $\sigma_L = 7.0 \pm 0.3\,\mu\mathrm{m}$ (corresponding to logarithmic distribution parameters $\langle \ln L \rangle = 2.60$ and $\sigma_{\ln L} = 0.44$).
    \item \textbf{Scale prefactor $\alpha$}: The dimensionless prefactor $\alpha$ relates the freeze-out correlation length to the defect density through $n\propto\alpha^d\bar{\xi}_f^{-d}$. The values $\alpha=4.0$, $6.9$, and $12.0$ illustrate the sensitivity of the absolute density scale to this nonuniversal normalization. In the comparison with experiment, $\alpha=12$ provides the closest agreement with the measured density scale and is therefore treated as an empirical normalization rather than an independent model prediction. Its value does not affect the crossover shape or apparent exponent.
\end{itemize}

Next, we evaluate the finite-size shift in the critical temperature, $\Delta T_C$ (Eq.~\ref{eq:crit_temp_shift}). Substituting the parameters $s_Q^z$ and $\tilde{a}$ obtained from density functional theory calculations for \ce{ErMnO3} \cite{meier2017global}:
\begin{align}
\Delta T_C &= \frac{\pi^2 \cdot s_Q^z \cdot T_{C0}}{\tilde{a} \cdot L_s^2} \nonumber \\
&= \frac{\pi^2 (18.9 \text{ eV}) (1429 \text{ K})}{(3.8 \text{ eV \AA}^{-2})(30 \times 10^4 \text{ \AA})^2} \nonumber \\
&= 7.8 \times 10^{-7} \text{ K}.
\label{eq:crit_temp_shift_param}
\end{align}
Because $\Delta T_C < 10^{-6}\text{ K}$ across all experimentally relevant grain sizes ($L_s \ge 1\,\mu\mathrm{m}$), finite-size shifts in $T_C$ are completely negligible and all subsystems undergo the transition at effectively the same bulk critical temperature.

\section{Saddle-point solution to 3D XY under Dirichlet boundary conditions}
\label{sec:TheoryExploreGP}

\subsection{Justifications for abrupt changes in correlation length}
\label{sec:GPsuperfluid}

Following the work of Ginzburg and Pitaevskii \cite{ginzburg1958theory}, the Ginzburg Landau (GL) $\psi^4$ thermodynamic potential of the superfluid phase transition in $^4$He may be written as 
\begin{align}
    f( \psi (\bm r)) = f_0 + \kappa |\nabla  \psi (\bm r) |^2 - \alpha | \psi (\bm r)|^2 + \frac{\beta}{2} | \psi (\bm r)|^4
\end{align}
where $\psi (\bm r)$ is the complex-valued field of the superfluid order parameter. The fourth-order GL free energy with respect to a complex order parameter, $\psi (\bm r) = \left( \beta / \alpha \right)^{1/2} \psi_0 (\bm r) \exp(i \phi (\bm r))$ can be restated as an isotropic two-component vector field. This makes apparent that the superfluid $^4$He phase transition falls within the same universality class as the classical XY model, as both can be described by the same field theory, lending themselves to the same RG analysis, and therefore an identical set of critical exponents.

Ginzburg and Pitaevskii assume a uniform phase in their derivation of $\psi_0$ subject to Dirichlet boundary conditions (DBCs). In other words, a mean-field approximation is made in order to examine the spatial variation of the magnitude of the order parameter, confined between two plates of thickness $L_s$. Due to the symmetry of the system, the steady-state differential equation can be cast into a one-dimensional ordinary differential equation, stated as Equation~\ref{eq:psi0_eqn_integrated}.

%
\begin{equation}
\begin{aligned}
\left| \frac{\partial \psi_0}{\partial z} \right|^2 +& \psi_0^2 - \frac{1}{2} \psi_0^4 = C \\
\quad \psi_0(z) \bigg\rvert_{z=0} =0 \quad \& & \quad \frac{\partial \psi_0 (z)}{\partial z} \bigg\rvert_{z=L_s/2} = 0.
\end{aligned}
\label{eq:psi0_eqn_integrated}
\end{equation}
The above boundary value problem, Equation~\ref{eq:psi0_eqn_integrated}, is normalized to be dimensionless, with the spatial coordinate rescaled as $\tilde z = z / l$, where $l = \left( \kappa / \alpha \right)^{1/2}$. In the Gaussian field theory model of critical fluctuations, $l$ corresponds to the correlation length \cite{kardarStatisticalPhysicsFields2007}. Perturbative techniques are needed to include the effects of terms that have higher order than quadratic, i.e. $\psi_0^4$ \cite{kardarStatisticalPhysicsFields2007}. Using the phenomenological relationship between $\alpha$ and temperature, $\alpha = \tilde \alpha \theta = \tilde \alpha (T_C - T) / T_C$, it is clear that $l$ approaches infinity at the critical point with $l \sim \theta^{-1/2}$ scaling behavior.

\begin{figure}[H]
    \centering
    \includegraphics[width=0.6\linewidth]{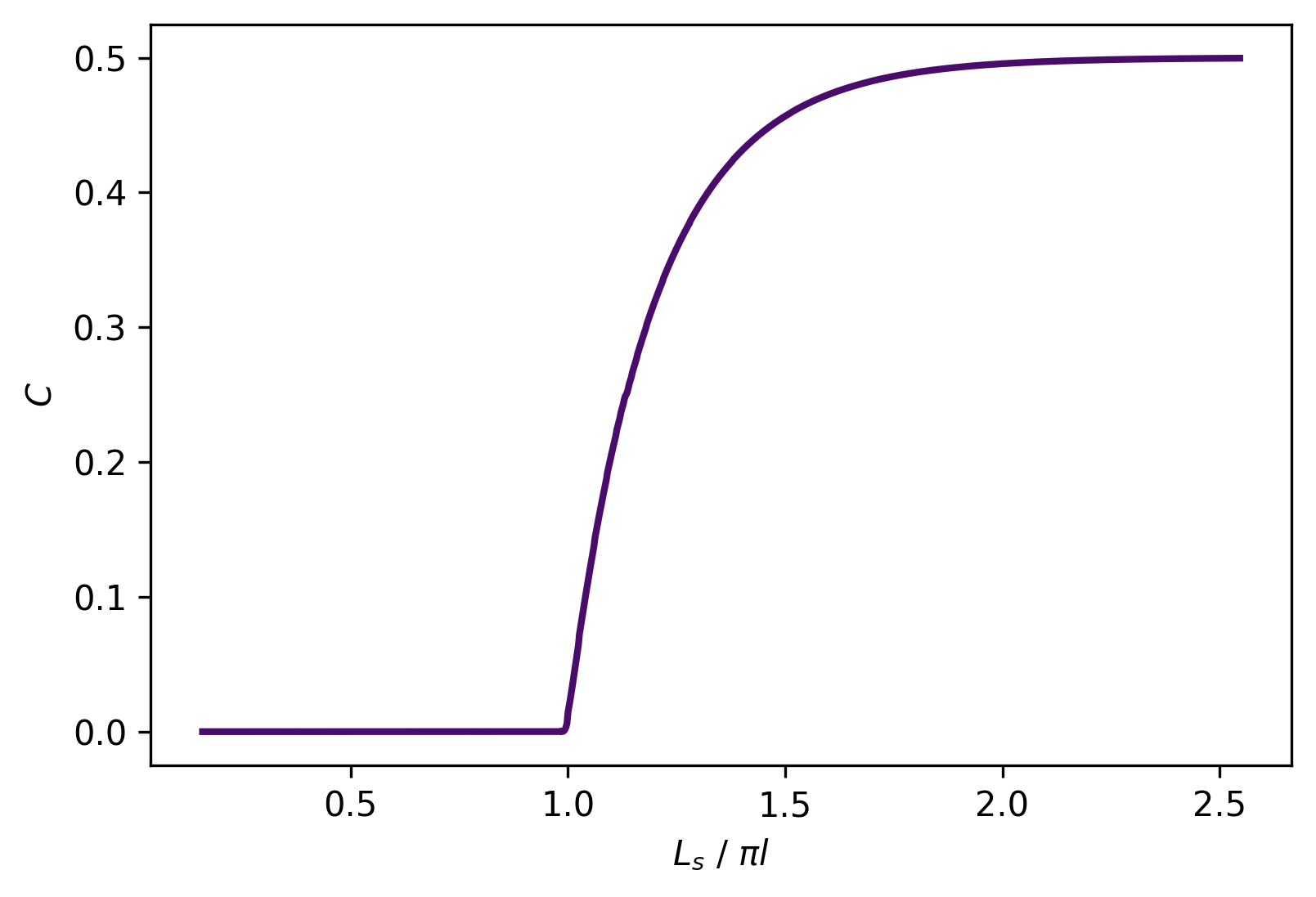}
    \caption{\textbf{The integration constant, $C$, for the free energy in Eq.~\ref{eq:psi0_eqn_integrated} versus the normalized system size $L_s$.} The discontinuity in the derivative of this curve at $L_s / \pi l = 1$ indicates a critical $L_s$, denoted as $\overline L_{s} = \pi l$. }
    \label{fig:critical_d}
\end{figure}

Eq.~\ref{eq:psi0_eqn_integrated} can be solved, following from Ref.~\cite{ginzburg1958theory}, as the following, 
\begin{equation}
\begin{aligned}
z &= \int_0^{\psi_0} \frac{d \psi_0}{\sqrt{\psi_0^4 / 2-\psi_0^2+C}}=\frac{\sqrt{2}}{b} \int_0^{\psi_0 / a} \frac{d u}{\sqrt{\left(1-u^2\right)\left(1-k^2 u^2\right)}} \\
& \equiv \frac{\sqrt{2}}{b} F(\varphi, k), \quad \varphi=\arcsin \left(\psi_0 / a\right), k=a / b, \\
& a^2=(1-\sqrt{1-2 C}), \quad b^2=(1+\sqrt{1-2 C)},
\end{aligned}
\label{eq:psi0_soln_dirichlet}
\end{equation}
where $F(\phi,k)$ is an elliptic integral, and the constant of integration, $C$, is obtained from solving the following equation,
\begin{align}
\frac{L_s}{2 l}=\frac{\sqrt{2}}{b} F\left(\frac{\pi}{2}, \frac{a}{b}\right)
\end{align}
which places the constraint on the midpoint value of the function governed by the DBCs and the mirror symmetry of the system at $z = L_s/2$. Ginzburg and Pitaevskii then showed that there is a critical thickness of the system, at which point the integration constant $C$ goes to zero with derivative discontinuity. The physical meaning of $C$ from Eq.~\ref{eq:psi0_eqn_integrated} can be thought of as a free energy of the system per unit volume, in reference to the $\psi_0 (z) = 0$ state.
Therefore, at the critical thickness, $\overline L_s$, the solution changes from zero at the end points, to a function that is zero everywhere on the subsystem.

In Fig.~\ref{fig:phi4_dirichlet}, we show the plot of the solution of the order parameter magnitude in space for various values of the integration constant $C$. Once again, the solution is ``mean-field'' in the sense that there's no variation of the phase of the order parameter. In the context of the classical XY model, this translates to the constraint that the magnetization direction is constant throughout the sample. Therefore, there is no consideration of vortex-vortex interactions. This simplification has obvious ramifications in the context of the hexagonal manganite system considered in this study, and furthermore regarding the link between the topological defect density and the correlation length. That being said, we explore the results from this seminal study by Ginzburg and Pitaevskii, because it serves to provide some justification for the sharp derivative of the effective correlation length versus temperature in our simple model: under DBCs, there is a critical thickness below which symmetry breaking Goldstone modes are suppressed by the boundary conditions.

\begin{figure}[H]
    \centering
    \includegraphics[width=0.6\linewidth]{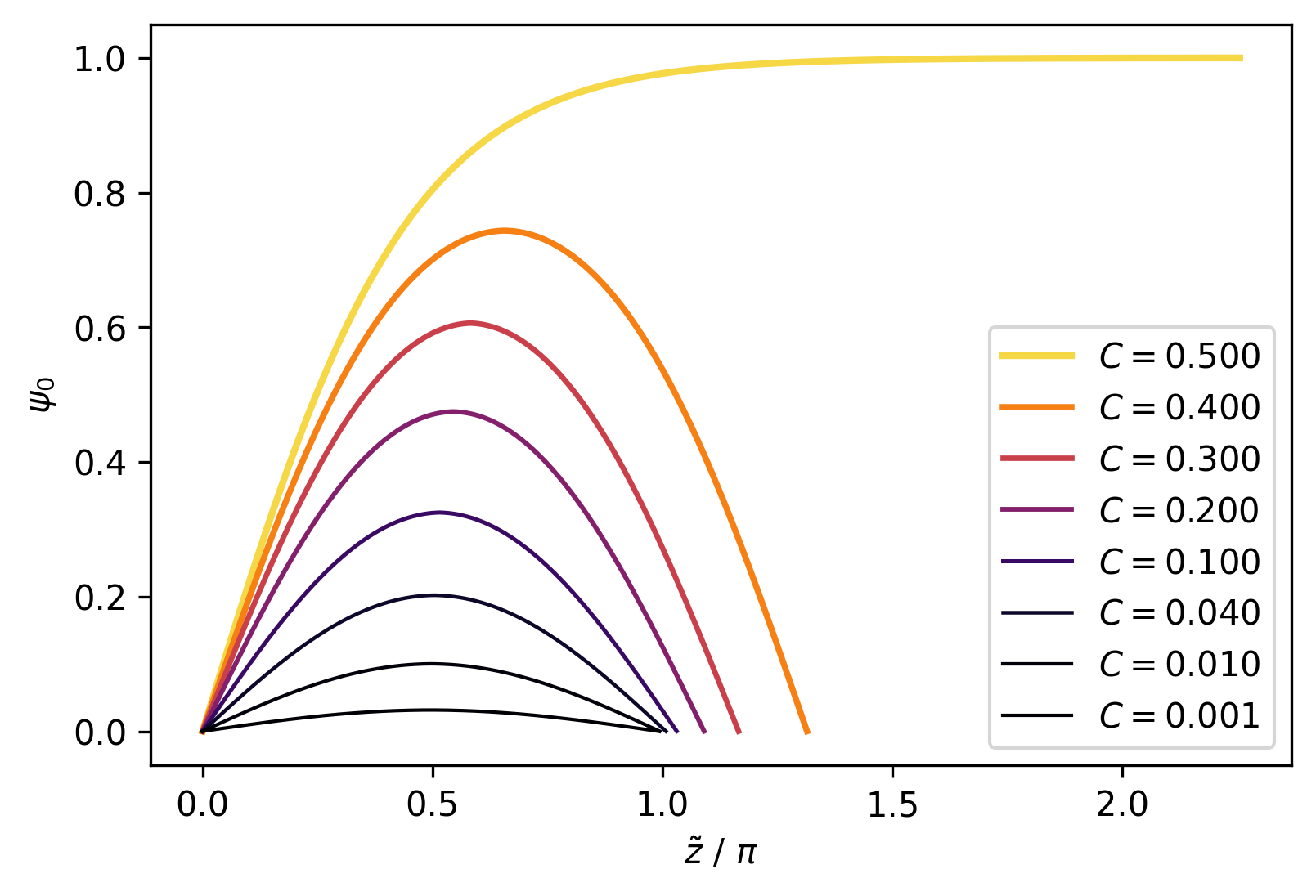}
    \caption{\textbf{Solution to Eq.~\ref{eq:psi0_soln_dirichlet}, plotted for various values of the integration constant $C$ from Eq.~\ref{eq:psi0_eqn_integrated}.} This provides a visual link between the shape of $\psi_0(z)$ versus the system size $L_s$ via the integration constant $C$, as shown in Fig.~\ref{fig:critical_d}. It is apparent that for $L_s$ smaller than the critical size $\overline L_s$, $\psi_0(z) = 0$ for all $z$.}
    \label{fig:phi4_dirichlet}
\end{figure}



\subsection{Justification for neglect of $\Delta T_c$}
\label{sec:DeltaTempFiniteSize}

Perhaps even more importantly, Ginzburg and Pitaevskii in Ref.~\cite{ginzburg1958theory} provide a justification for our neglect of the effect of finite size effects on the change in the critical temperature. A system of size $L_s$ has a critical temperature $T_C$ shifted by
\begin{align}
  \Delta T_C &= \frac{\pi^2 \cdot \kappa \cdot T_{C0}}{\tilde{\alpha} \cdot L_s^2}
  \label{eq:crit_temp_shift}
\end{align}
compared to the bulk limit, $L_s \rightarrow \infty$. A detailed derivation of this equation in Ref.~\cite{ginzburg1958theory}, but it can be solved by substituting $l$ with $L_s / \pi$ in the relationship between $l$ and $T_C$.

If $\Delta T_C$ were sufficiently sensitive to $L_s$, then the model developed in Sec.~\ref{sec:math_details} would need to be modified to include a statistical average over $T_C$. But, reassuringly, in Sec.~\ref{sec:parameters}, from Eq.~\ref{eq:crit_temp_shift_param}, we show that $\Delta T_C < 10^{-6} K$, and therefore the finite-size affect on $T_C$ can be neglected entirely from our analysis.

\section{Experimental details}
\label{sec:AddExpDetails}

\subsection{Cooling protocol}
\label{sec:CoolingProtocol}
As displayed in Fig.~\ref{fig:figS1}, pressed \ce{ErMnO3} powder pellets were sintered at 1750\,K for 24\,h to ensure that the microstructure remains largely unaffected by the different cooling rates. Cooling over $T_C$=1430\,K was done with varying cooling rates of 10$^{-2}$\,K\,min$^{-1}$, 10$^{-1}$\,K\,min$^{-1}$, 10$^{0}$\,K\,min$^{-1}$, and 10$^{1}$\,K\,min$^{-1}$ in a temperature range 20\,K above and below of \textit{T$_{c}$} as indicated by the different colors. Outside the 20\,K range a cooling rate of 5\,K\,min$^{-1}$ was utilized.
\begin{figure}[H]
    \centering
    \includegraphics[width=0.4\linewidth]{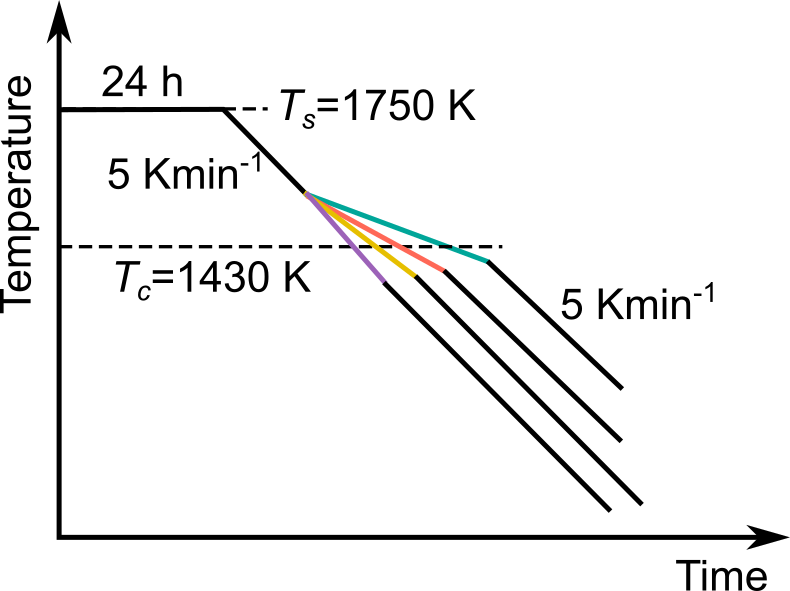}
    \caption{\textbf{Cooling profile.} Cooling through $T_c = 1430$\,K was carried out at $10^{-2}$ to $10^{1}$\,Kmin$^{-1}$ within $\pm$20\,K of $T_c$ (colors), and at 5\,Kmin$^{-1}$ otherwise.}
    \label{fig:figS1}
\end{figure}

\subsection{Crystal- and microstructure}
\label{sec:Microstructure}
To verify that variations in cooling rate do not alter the crystal structure or microstructure, we characterized samples cooled under representative slow and fast quench conditions. As shown in Fig.~\ref{fig:figS1a}, X-ray diffraction and microscopy confirm phase purity and comparable grain morphology across all cooling rates, ensuring that the observed differences in vortex density arise from nonequilibrium dynamics rather than structural or microstructural changes.
\begin{figure}[H]
    \centering
    \includegraphics[width=0.7\linewidth]{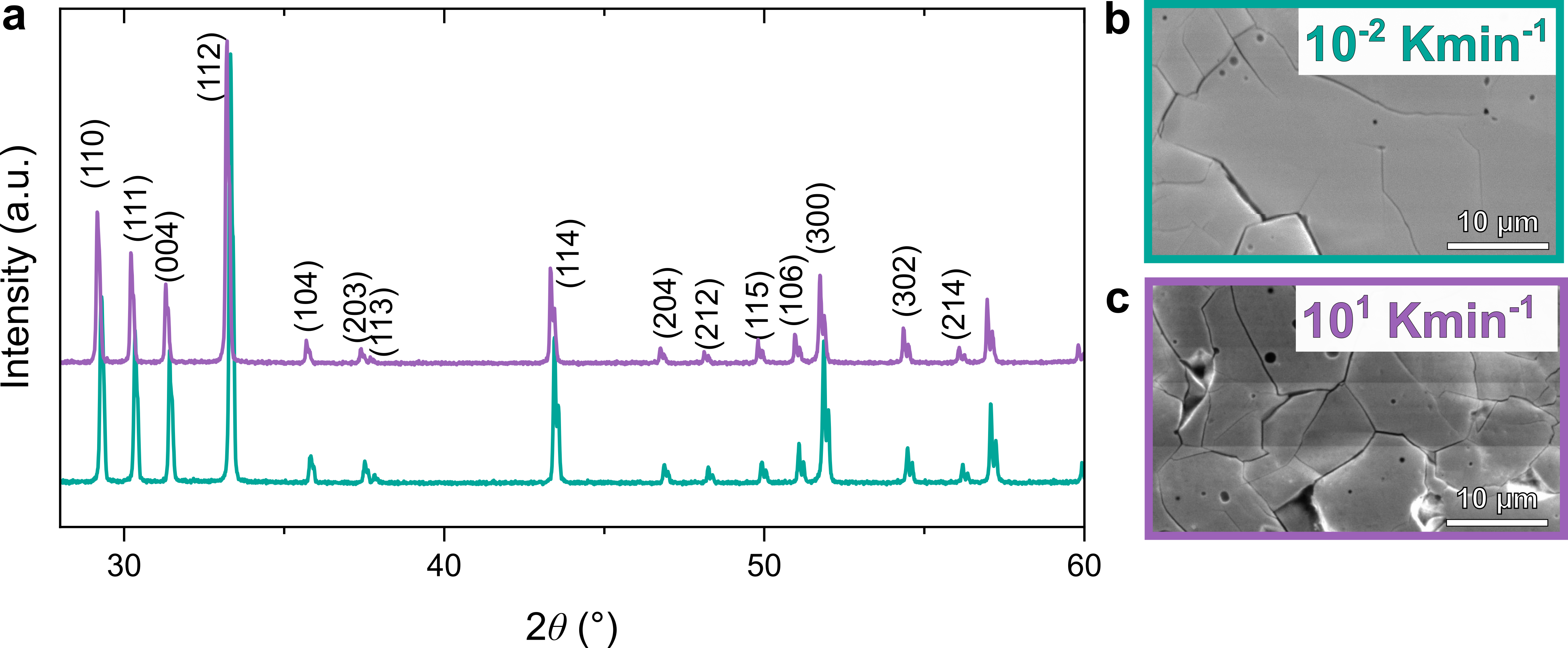}
    \caption{\textbf{Crystal- and microstructure of samples cooled at different rates.} \textbf{a)} XRD patterns of polycrystalline \ce{ErMnO3} cooled under 10$^{-2}$\,K\,min$^{-1}$ and 10$^{1}$\,K\,min$^{-1}$ with peaks of the space group $P6_{3}cm$ marked. The micrographs of the samples are displayed in \textbf{(b)} and \textbf{(c)}. For all cooling rates the grain size, and the distribution of grain size is approximately independent of the cooling rate.}
    \label{fig:figS1a}
\end{figure}

\subsection{Domain morphology under different cooling rates}
\label{sec:PeriodicityDomain}
To quantify the influence of cooling rate on domain morphology, we analyze the evolution of stripe periodicity across a wide range of quench conditions. As shown in Fig.~\ref{fig:figS2}, the average stripe spacing, extracted from 80 positions per sample, systematically varies with cooling rate, consistent with the representative PFM images displayed.
\begin{figure}[H]
    \centering
    \includegraphics[width=0.6\linewidth]{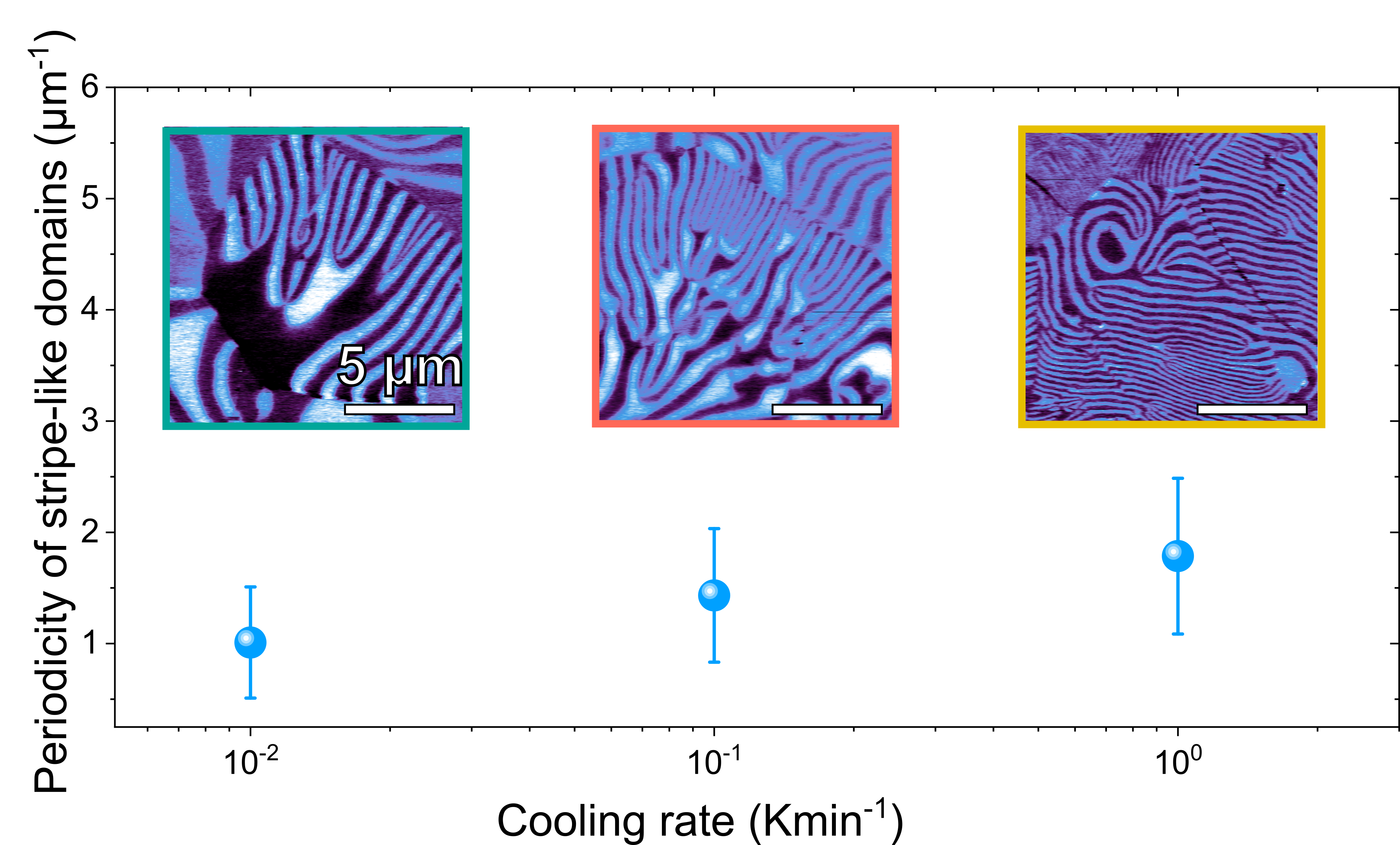}
    \caption{\textbf{Domain morphology under different cooling rates.} The periodicity of stripe-like domains averaged over 80 positions is displayed as a function of the cooling rate. Representative PFM images of samples cooled under different rates are displayed.}
    \label{fig:figS2}
\end{figure}

\section{Effect of strain-induced anisotropy}
\label{sec:StrainAnisotropy}

\begin{figure}[H]
    \centering
    \includegraphics[width=0.95\linewidth]{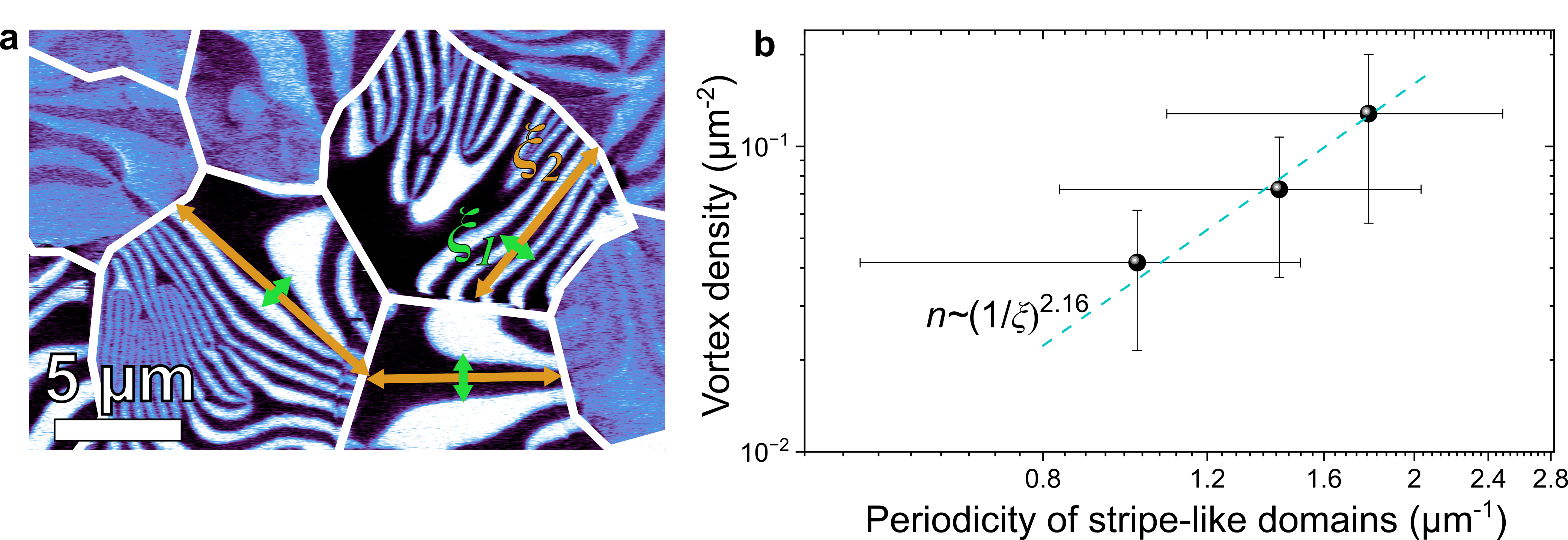}
    \caption{\textbf{Influence of strain on the vortex density.} \textbf{(a)} Representative PFM data displaying stripe-like ferroelectric domain structures with two correlation-lengths, $\xi_1$ , and $\xi_2$. One of the correlation lengths, $\xi_2$, is truncated by the grain size. \textbf{(b)} Plot of vortex density (obtained from Fig.~\refM{fig:combined}{3}e) as a function of the inverse periodicity of the stripe-like domains (obtained from Fig.~\ref{fig:figS2}). The data can be described by a power function with an exponent of $2.16 \pm 0.34$, indicating that the validity of the relationship between the vortex density and a single correlation length, $n \propto \xi^{-2}$.}
    \label{fig:figS3}
\end{figure}

Understanding the interplay between strain-induced symmetry breaking and correlation length is crucial to elucidate the KZ scaling behavior of our polycrystalline materials. In polycrystalline hexagonal manganites, the non-uniform orientation of the anisotropic grains introduces an inhomogeneous strain,~\cite{Tomczyk2012} which profoundly impacts their ferroelectric domain structure.~\cite{schultheissConfinementDrivenInverseDomain2022} Previous phase field simulations have shed light on this relationship, particularly highlighting the stripe-like ``incommensurate" domain structure as a direct consequence of these strain fields \cite{schultheissConfinementDrivenInverseDomain2022, Xue2017}. Central to our investigation on the influence of size effects on the critical exponent of the KZM is the exploration of the scaling relationship, $n = \xi^{-2}$. As displayed in Fig.~\ref{fig:figS3}a, the strain component in our polycrystals induces symmetry breaking, which yields two correlation-lengths, $\xi_1$, and $\xi_2$. Thereby, $\xi_1$ is the incommensurate characteristic length, and $\xi_2$ is the longitudinal correlation length perpendicular to the incommensurate wave vector, such that $\xi_1 < \xi_2$. In this setting, the vortex density should scale as $n = (\xi_1 \times \xi_2)^{-1}$, where a singularity of $\xi_2$ at critically is suppressed by the grain size. 
We propose a scenario, wherein for $\xi_2$ exceeding the grain size, $L$, $\xi_2$ is truncated such that $\xi_2=L$, leading to a scaling behavior $n \propto (\xi_1 \times L)^{-1} \propto \xi_1^{-1}$. This would manifest as a dimensionality reduction in the expression for $\mu$, explicitly $\mu = \frac{\nu}{1 + z\nu} d$ (exponent in Eq.~\refM{eq:KZMexponent}{1}), where $d$ is the dimension of the system . Therefore, the reduction in $d$ from $d=2$ (single crystal) to $d=1$ (polycrystal) could explain the experimentally observed reduction from $\mu=0.58$ to $\mu=0.25$ (Fig.~\refM{fig:combined}{3}d).
Utilizing data from our polycrystalline \ce{ErMnO3} samples cooled under different rates, we, however, can ascertain the validity of the scaling relationship, $n = \xi^{-2}$. Experimental data of vortex density and correlation length determined at different cooling rates corroborate our hypothesis, with a fit of a power law (Fig.~\ref{fig:figS3}b) yielding an exponent of $2.16 \pm 0.34$, confirming $n \propto \xi^{-2}$. Consequently, we can conclude that the observed reduction in $\mu$ primarily arises from finite-size scaling, while strain effects resulting in a dimensionality reduction may play a secondary role.

\section{Effect of Thermalization}
\label{sec:ThermalizationEffects}

\begin{figure}[H]
    \centering
    \includegraphics[width=0.95\linewidth]{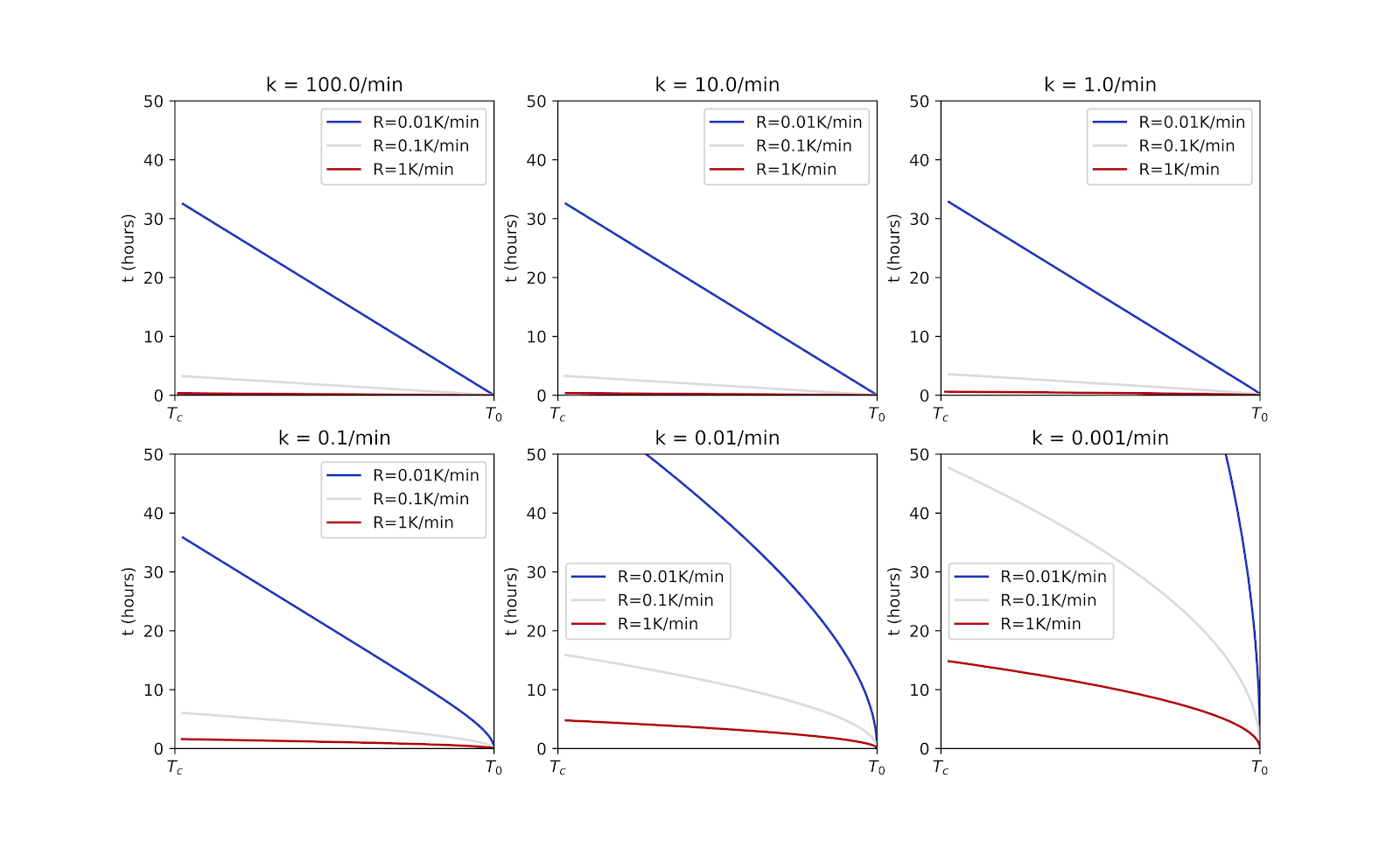}
    \caption{\textbf{Influence of finite cooling dynamics on the effective quench rate.} A series of temperature profiles $T(t)= T_0 - Rt + \frac{R(1-e^{-kt})}{k}$ of different $R$ and $k$ choices, illustrating that for the temperature and quench rate range used in this work, $k$ need be on the order of roughly 1/min or below to have an effect on the effective quench rate.}
    \label{fig:figS4s}
\end{figure}

While the samples shown here were subjected to a thermal bath that was annealed at a linear rate, the effective rate of change of the sample temperature is not necessarily linear. This is due to the fact that the sample temperature will equilibrate exponentially to that of the bath, so the true sample temperature can be obtained from the convolution of these two effects. From Newton's law of cooling, we then have the following differential equation representing the instantaneous temperature of the sample $T_s(t)$ in terms of the heat transfer coefficient $k$ and the bath temperature $T_b(t)$. 
\begin{align}
\frac{dT_s}{dt} = - k (T_s(t) - T_b(t))
\end{align}
Using a power series solution $T_s(t)=\sum_n c_n t^n$ and taking $T_b(t) = T_0 - Rt$, we find the following recursion relation with $c_0=T_0$.
\begin{align}
c_{n+1} = \frac{k}{n + 1} ( c_0 \delta_{n,0} - R\delta_{n,1} - c_n )
\end{align}
Which leads to the following. 
\begin{align}
\label{eq:33}
T_s(t)= T_0 + \sum_{n=2} \frac{Rt^{n}}{n!} (-k)^{n-1} 
=  T_0 - Rt + \frac{R(1-e^{-kt})}{k}
\end{align}
We can see from this result that the sample temperature does not change linearly even for this imposed linear quench so long as the equilibration rate is not exceptionally large compared to the quench rate. This can be seen by noting that samples with larger conductivities will instead equilibrate instantaneously on the time scale associated with the quench and therefore display the same linear cooling as the bath (Eq.~\ref{eq:33} converges to $T_0 - Rt$ in the limit of large $k$), as is commonly assumed in most studies of the KZ mechanism we are aware of, apart from work investigating weak thermal coupling in the context of the quantum KZ mechanism \cite{nalbach2015quantum}. We also note that most work developing the standard framework for finite system sizes was performed using the gradual removal of external fields \cite{Binder1987} in which a consideration of conductivity is not necessary. 

To demonstrate that this effect cannot explain the trends shown, we obtain a sensible value for the proportionality constant $k$ from measured quantities. We can assume that heat transfer between the bath and the sample proceeds primarily through the $c$-axis, motivated by the higher $c$-axis thermal conductivity shown for \ce{ErMnO3} and sample geometry.~\cite{pang2023thermal} For an approximate treatment of heat transfer, we treat the center of the sample as $T_s(t)$ and each interface above and below as $T_b(t)$ and consider equilibriation between $T_s$ and only one of the two fronts. This yields a rule of thumb estimate of $k = \frac{2\kappa_c A_{ab}}{C_p L_c}$ where $\kappa_c$ and $C_p$ are the $c$-axis thermal conductivity and the (constant pressure) heat capacity, respectively. A more thorough consideration of heat transfer between both interfaces is not needed for this very qualitative treatment, as we primarily aim to show that the order of magnitude for this effect is too subtle to explain the trend seen. Within the temperature range investigated experimentally (1430-1750 K), the $C_p$ of \ce{ErMnO3} is in the range of roughly $(155 \pm 5)\,\mathrm{J\,mol^{-1}\,K^{-1}}$ (extrapolated from measurements up to 900\,K~\cite{satoh1995high}). The out of plane thermal conductivity $\kappa_c$ is weakly dependent on temperature (at least in the 0-300\,K range) and on the order of $5\text{--}10\,\mathrm{W\,m^{-1}\,K^{-1}}$.~\cite{pang2023thermal} We note that prior work has also demonstrated that $\kappa_c$ depends on the ferroelectric domain size.~\cite{pang2023thermal,belrhiti2025domain} However within the range of vortex densities measured here of roughly $0.01\text{--}0.1\,\mathrm{\mu m^{-2}}$, the values of $\kappa_c$ within single crystal samples fell within the narrow range of roughly $5\text{--}7\,\mathrm{W\,m^{-1}\,K^{-1}}$. This yields a $k$ on the order of $1000\,\mathrm{min^{-1}}$ when $L_c \approx 1\,\mathrm{mm}$, $A_{ab} \approx 300\,\mathrm{mm^2}$, $C_p \approx 155\,\mathrm{J\,mol^{-1}\,K^{-1}}$, and $\kappa_c \approx 7\,\mathrm{W\,m^{-1}\,K^{-1}}$. The measured thermal conductivity of the polycrystalline samples is smaller by approximately $50\%$ and its morphology is comparable.~\cite{belrhiti2025domain} We can see in Fig.~\ref{fig:figS4s} for the sample conductivity to have an impact on the effective thermal quench rate, the coefficient $k$ must be on the order of at least $1\,\mathrm{min^{-1}}$. This would require thermal conductivities roughly 1000 times smaller or sample thicknesses 30 times larger than those used here, motivating our attribution of the modified critical scaling to other origins.

\end{widetext}
  \else
  \fi
\fi


\end{document}